\documentclass[twocolumn]{bmcart}% uncomment this for twocolumn layout and comment line below
\usepackage{amsthm,amsmath}
\usepackage[utf8]{inputenc} %unicode support
\usepackage{graphicx}
\usepackage{amsmath, amssymb}
\usepackage[printonlyused,nolist]{acronym}
\usepackage{xcolor}
\usepackage{algorithm}
\usepackage{algpseudocode}
\usepackage{etoolbox}
\usepackage{tikz}
\usetikzlibrary{calc}
\usepackage{url}
\usepackage{multirow}
\usepackage{tikz}
\usepackage{pgfplots}
\usepackage{optidef}
\usepackage{booktabs}
\usepackage{gensymb}
\usepackage{LMSnodeshapes}
\usepackage{pifont}
\usepackage{slashbox}
\usepackage{xcolor}
\usepackage{algorithm}
\usepackage{algpseudocode}
\begin{acronym}
	\acro{ENR}{Echo-to-Noise Ratio}
	\acro{KF}{Kalman Filter}
	\acro{PBAF}{Partitioned-Block Adaptive Filter}
	\acro{ML}{Maximum-Likelihood}
	\acro{EM}{Expectation-Maximization}
	\acro{AEC}{acoustic echo cancellation}
	\acro{PSD}{power spectral density}
	\acro{ERLE}{Echo Return Loss Enhancement}
	\acro{AIR}{acoustic impulse response}
	\acro{ATF}{acoustic transfer function}
	\acro{FIR}{Finite Impulse Response}
	\acro{NLMS}{normalized least-mean-squares}
	\acro{NMF}{Nonnegative Matrix Factorization}
	\acro{PDF}{Probability Density Function}
	\acro{VSS}{Variable Step Size}
	\acro{DFT}{Discrete Fourier Transform}
	\acro{NER}{Near-End-to-Echo-Ratio}
	\acro{ENR}{Echo-to-Noise-Ratio}
	\acro{SISO}{single-input single-output}
	\acro{SIMO}{single-input multiple-output}
	\acro{MISO}{multiple-input single-output}
	\acro{MIMO}{multiple-input multiple-output}
	\acro{AF}{adaptive filter}
	\acro{WGN}{white Gaussian noise}
	\acro{FDAF}{Frequency-Domain Adaptive Filter}
	\acro{SNR}{Signal-to-Noise-Ratio}
	\acro{SSFDAF}{State-Space Frequency-Domain Adaptive Filter}
	\acro{MCSSFDAF}{Multichannel State-Space Frequency-Domain Adaptive Filter}
	\acro{RTF}{Real-Time Factor}
	\acro{NN}{nearest neighbour}
	\acro{KNN}{k-nearest neighbour}
	\acro{PCA}{Principal Component Analysis}
	\acro{KF-ASP}{KF with an Adaptive Subspace Projection}
	\acro{OSASI}{Online Supervised Acoustic System Identification}
	\acro{MMSE}{minimum mean square error}
\end{acronym}

\newcommand{\commentTHa}[1]{\textcolor{black}{#1}}
\newcommand{\commentTHb}[1]{\textcolor{black}{#1}}
\newcommand{\commentTHc}[1]{\textcolor{black}{#1}}
\newcommand{\commentTHd}[1]{\textcolor{black}{#1}}
\newcommand{\commentTHe}[1]{\textcolor{black}{#1}}

\startlocaldefs
\endlocaldefs

\DeclareMathOperator*{\w}{\boldsymbol{w}}
\DeclareMathOperator*{\X}{\boldsymbol{X}}
\DeclareMathOperator*{\bP}{\boldsymbol{P}}
\DeclareMathOperator*{\e}{\boldsymbol{e}}
\DeclareMathOperator*{\C}{\boldsymbol{C}}
\DeclareMathOperator*{\y}{\boldsymbol{y}}

\DeclareMathOperator*{\herm}{\text{H}}
\DeclareMathOperator*{\trans}{\text{T}}

\newcommand{\kfMean}{{\hat{\boldsymbol{w}}}^{\text{kf}}}

\begin{document}

%%% Start of article front matter
\begin{frontmatter}

\begin{fmbox}
\dochead{Research}


% \title{Noise-Robust Acoustic System Identification Exploiting Kalman Filtering and an Adaptive System Subspace Model}
\title{Online Acoustic System Identification Exploiting Kalman Filtering and an Adaptive Impulse Response Subspace Model}

%%%%%%%%%%%%%%%%%%%%%%%%%%%%%%%%%%%%%%%%%%%%%%
%%                                          %%
%% Enter the authors here                   %%
%%                                          %%
%% Specify information, if available,       %%
%% in the form:                             %%
%%   <key>={<id1>,<id2>}                    %%
%%   <key>=                                 %%
%% Comment or delete the keys which are     %%
%% not used. Repeat \author command as much %%
%% as required.                             %%
%%                                          %%
%%%%%%%%%%%%%%%%%%%%%%%%%%%%%%%%%%%%%%%%%%%%%%

\author[
  addressref={aff1},                   % id's of addresses, e.g. {aff1,aff2}
  corref={aff1},                       % id of corresponding address, if any
% noteref={n1},                        % id's of article notes, if any
  email={thomas.haubner@fau.de}   % email address
]{\inits{T.} \fnm{Thomas} \snm{Haubner}}
\author[
  addressref={aff1},
  email={andreas.brendel@fau.de}
]{\inits{A.}\fnm{Andreas} \snm{Brendel}}
\author[
addressref={aff1},
email={walter.kellermann@fau.de}
]{\inits{W.}\fnm{Walter} \snm{Kellermann}}

%%%%%%%%%%%%%%%%%%%%%%%%%%%%%%%%%%%%%%%%%%%%%%
%%                                          %%
%% Enter the authors' addresses here        %%
%%                                          %%
%% Repeat \address commands as much as      %%
%% required.                                %%
%%                                          %%
%%%%%%%%%%%%%%%%%%%%%%%%%%%%%%%%%%%%%%%%%%%%%%

\address[id=aff1]{%                           % unique id
  \orgdiv{Multimedia Communications and Signal Processing},             % department, if any
  \orgname{Friedrich-Alexander-University Erlangen-Nürnberg (FAU)},          % university, etc
  \city{Cauerstr. 7, 91058, Erlangen},                              % city
  \cny{Germany.}                                    % country
}

%%%%%%%%%%%%%%%%%%%%%%%%%%%%%%%%%%%%%%%%%%%%%%
%%                                          %%
%% Enter short notes here                   %%
%%                                          %%
%% Short notes will be after addresses      %%
%% on first page.                           %%
%%                                          %%
%%%%%%%%%%%%%%%%%%%%%%%%%%%%%%%%%%%%%%%%%%%%%%

%%\note{Sample of title note}     % note to the article
%\note[id=n1]{Equal contributor} % note, connected to author

% \end{fmbox}% comment this for two column layout

%%%%%%%%%%%%%%%%%%%%%%%%%%%%%%%%%%%%%%%%%%%%%%%
%%                                           %%
%% The Abstract begins here                  %%
%%                                           %%
%% Please refer to the Instructions for      %%
%% authors on https://www.biomedcentral.com/ %%
%% and include the section headings          %%
%% accordingly for your article type.        %%
%%                                           %%
%%%%%%%%%%%%%%%%%%%%%%%%%%%%%%%%%%%%%%%%%%%%%%%

\begin{abstractbox}

\begin{abstract} % abstract
%\justify
We introduce a novel algorithm for online estimation of acoustic impulse responses (AIRs) which allows for \commentTHd{fast} convergence by exploiting prior knowledge about the \commentTHd{fundamental structure} of AIRs. The proposed method assumes that the variability of AIRs of an acoustic scene is confined to a low-dimensional manifold which is embedded in \commentTHd{a} high-dimensional space of \commentTHd{possible} AIR estimates. We discuss various approaches to locally approximate the AIR manifold by affine subspaces which are assumed to be tangential hyperplanes to the manifold. The validity of \commentTHd{these} model assumptions is \commentTHd{verified} for simulated data.
% The respective parameters of the affine subspaces are learned from training data.
% The manifold is locally approximated by an affine subspace whose parameters are computed from training data. 
Subsequently, we describe how the learned models can be used to \commentTHc{improve} online AIR estimates by projecting them onto an adaptively estimated subspace. The \commentTHd{parameters determining the subspace are learned from training samples in a local neighbourhood to the current AIR estimate.} This allows the system identification algorithm to benefit from \commentTHd{preceding estimates in the acoustic scene}. 
% As the subspace parameters are learned from training data, the proposed system identification algorithm can benefit from preceding estimates in similar acoustic environments. 
% which is learned from training data. 
% This allows the online system identification algorithm to benefit from preceding estimates in similar acoustic environments. 
% The respective parameters of the subspace are learned from the closest training data samples to the current AIR estimate. 
% This allows the online system identification algorithm to benefit from preceding estimates in similar acoustic environments. 
To assess the proximity of training data AIRs \commentTHc{to the current {AIR} estimate}, we introduce a probabilistic extension of the Euclidean distance which improves the performance for applications with correlated excitation signals. Furthermore, we describe how model imperfections can be tackled by a \commentTHd{soft} projection of the AIR estimates.
% a relaxed projection for denoising the AIR estimates to deal with imperfections of the manifold model. 
The proposed algorithm exhibits significantly faster convergence properties in comparison to a high-performance state-of-the-art algorithm. \commentTHc{Furthermore, we show an improved steady-state performance for speech-excited system identification scenarios suffering from high-level interfering noise and nonunique solutions.}
% Furthermore, we show an improved steady-state system identification performance for scenarios suffering from high-level interfering noise and non-unique system excitation. %\textcolor{red}{(In the final version the abstract should be in 2-column style.)}

\end{abstract}

%%%%%%%%%%%%%%%%%%%%%%%%%%%%%%%%%%%%%%%%%%%%%%
%%                                          %%
%% The keywords begin here                  %%
%%                                          %%
%% Put each keyword in separate \kwd{}.     %%
%%                                          %%
%%%%%%%%%%%%%%%%%%%%%%%%%%%%%%%%%%%%%%%%%%%%%%


\end{abstractbox}
\end{fmbox}% uncomment this for two column layout

\end{frontmatter}

%%%%%%%%%%%%%%%%%%%%%%%%%%%%%%%%%%%%%%%%%%%%%%%%
%%                                            %%
%% The Main Body begins here                  %%
%%                                            %%
%% Please refer to the instructions for       %%
%% authors on:                                %%
%% https://www.biomedcentral.com/getpublished %%
%% and include the section headings           %%
%% accordingly for your article type.         %%
%%                                            %%
%% See the Results and Discussion section     %%
%% for details on how to create sub-sections  %%
%%                                            %%
%% use \cite{...} to cite references          %%
%%  \cite{koon} and                           %%
%%  \cite{oreg,khar,zvai,xjon,schn,pond}      %%
%%                                            %%
%%%%%%%%%%%%%%%%%%%%%%%%%%%%%%%%%%%%%%%%%%%%%%%%

%%%%%%%%%%%%%%%%%%%%%%%%% start of article main body
% <put your article body there>
\section{Introduction}
% \justify
\label{sec:intro}
The \commentTHa{continuously} increasing amount of acoustic communication devices has fueled the research on reliable speech enhancement algorithms. 
In this context system identification has proven to be \commentTHa{a} vital part of many state-of-the-art approaches \commentTHc{\cite{enzner_acoustic_2014, diniz_adaptive_filtering}}. In particular online algorithms are required to cope with the large \commentTHa{variety} of acoustic environments devices are exposed to. \commentTHc{However, even after decades of research \cite{widrow_b_adaptive_1960, ferrara_fast_1980, benesty-vss-lms, kuech_state-space_2014}, \ac{OSASI}-based speech enhancement algorithms are significantly challenged by interfering noise signals and their limited convergence rate.} %leading to an unsatisfactory user experience under such conditions.} 
%However, even after decades of research \commentTHa{\cite{widrow_b_adaptive_1960, ferrara_fast_1980, benesty-vss-lms, kuech_state-space_2014}} \ac{OSASI} algorithms have still not achieved a performance which \commentTHa{leads} to a satisfactory user experience. 
%
% This is mainly due to their susceptibility to interfering noise signals and their limited performance during the initial convergence phase. 
\commentTHc{In this} paper we propose a method which \commentTHc{tackles remaining} limitations of modern \ac{OSASI} algorithms.

Both convergence speed and noise-robustness are usually \commentTHc{addressed} \commentTHa{by adaptive step size-controlled \ac{AF} algorithms} \cite{diniz_adaptive_filtering}. \commentTHb{Their performance decisively depends on the stochastic properties of the excitation and the noise \commentTHd{signals} \cite{haykin_2002}. \commentTHe{In particular}, for stationary white excitation signals a fast convergence speed and robust steady-state performance is achieved.}
This observation has \commentTHa{led} to a variety of excitation signal-dependent adaptive step size \commentTHa{selection schemes} with the most famous one being the power-normalization of the \commentTHa{time-domain} \ac{NLMS} algorithm \cite{haykin_2002}. \commentTHe{Its scalar time-domain step size} has been extended to a frequency-dependent step size to cope with the temporal correlation of many excitation signals, i.e., speech or music \commentTHa{\cite{haensler2004acoustic}}. In particular the frequency-dependent power normalization in block processing approaches \commentTHa{led} to computationally efficient and faster converging algorithms \cite{ferrara_fast_1980, mansour_unconstrained_1982}. The robustness against nonstationary \commentTHd{interfering} noise sources was initially addressed by binary adaptation control, i.e., \commentTHc{stalling} the filter adaptation whenever the noise power \commentTHd{exceeds a predefined} threshold \cite{Benesty_new_2000}. % However, as binary adaptation control does not allow for continuous optimization, their tracking is limited. 
The scalar and binary decision was later extended to a frequency-dependent continuous step size control \cite{nitsch2000frequency}. In particular the probabilistic inference of the step size by a \ac{KF} has shown great potential \cite{enzner_frequency-domain_2006, malik_online_2010, 7115075}. Yet, the \commentTHa{\ac{KF}} performance decisively depends on an accurate estimate of the noise \ac{PSD} \cite{yang_frequency-domain_2017}. Here, significant performance improvements \commentTHa{relative to} classical \commentTHe{\ac{PSD}} estimators have been achieved by modern \commentTHa{machine-learning} based approaches \cite{kfNMF, haubner2020synergistic}. Despite the improved noise robustness, these approaches still achieve only slow convergence \commentTHa{speed} for scenarios suffering from \commentTHd{permanently} low \ac{SNR}, e.g., \commentTHd{as in} driving cars with open windows.

\commentTHc{Recently, besides} adaptation control, the exploitation of prior knowledge about the \commentTHa{structure} of \acp{AIR} has been successfully used to deal with \commentTHd{slow convergence and non-robust steady-state performance} \cite{fozunbal_multi-channel_2008, koren_supervised_2012, talmon_relative_2013, 9231543}. These algorithms rely on the assumption that not all \commentTHd{possible \ac{AIR} estimates, i.e., \ac{FIR} filters of fixed length,} are equally likely, i.e., certain regions exhibit a higher probability \commentTHa{of representing a valid \ac{AIR}}. \commentTHb{In \cite{fozunbal_multi-channel_2008} this assumption has been used by regularizing a least-squares system identification cost function with the Mahalanobis distance based on an estimated \ac{AIR} covariance matrix.}
%As this approach does also assume a single global \ac{AIR} covariance matrix it exhibits the same limitations as the single affine subspace approach \commentTHa{(cf.~Sec.~\ref{sec:mix_mode})}.}
The extreme case that \commentTHc{the \acp{AIR} of a considered acoustic scene can all be represented by a} structured subset of the high-dimensional estimation space\commentTHd{, i.e., \ac{FIR} filters of fixed length,} motivates the assumption of a low-dimensional \commentTHc{\ac{AIR}} manifold \cite{laufer-goldshtein_study_2015}. Its existence is often tightly coupled to the parameter changes of an underlying physical process, e.g., \commentTHc{location of sources and sensors or temperature changes, which govern} the variability of the \acp{AIR} \cite{talmon_diffusion_2013}. 
% \commentTHc{If it is possible to project noisy \ac{AIR} onto the manifold, noisy \ac{AIR} estimates}
\commentTHc{Noisy \ac{AIR} estimates can be enhanced by projection onto the manifold, i.e., by removing the part which is not confined to the manifold.}
%If it is possible to project \acp{AIR} estimates onto the manifold, the noisy estimates can be enhanced by \commentTHc{removing} the part which is not confined to the manifold.} 
%Yet it is not clear how a projection can be achieved. 
\commentTHa{Yet due} to the complex interaction of the physical parameters and the high-dimensional \acp{AIR}, \commentTHb{an analytic} manifold description is difficult to obtain. However, in many applications \commentTHa{a device is exposed to \commentTHe{a reoccurring acoustic scene}} 
%one faces reoccurring similar acoustic \commentTHa{scenes} 
which allows to \commentTHe{collect \ac{AIR} estimates. These estimates can serve as training data to optimize a data-driven AIR manifold model which \ac{OSASI} algorithms can exploit for improved performance.} 
% . This allows the algorithms to benefit from preceding estimates in \commentTHe{the acoustic environment.}
% Thus, the projection allows these methods to benefit from previous estimates obtained in a similar acoustic environment. 
% If the manifold can be learned from training data, noisy estimates can be enhanced by projection onto the manifold \cite{bibid}. 
Various approaches have been proposed to model an \ac{AIR} manifold with the most prominent one being a global affine subspace whose parameters are estimated by \ac{PCA} \cite{jolliffe1986principal}. However, \commentTHc{due to} the restrictive assumption that all \acp{AIR} are confined to a single affine subspace, this approach is limited to simplistic scenarios \commentTHe{(cf.~Sec.~\ref{sec:air_mod_par_est})}. In \cite{koren_supervised_2012, 9231543} this model has been extended to a mixture of affine subspaces  whose parameters are learned in advance by clustering \commentTHa{the \acp{AIR}} followed by \commentTHe{a \ac{PCA} of each cluster.} 
Due to the increased modelling capabilities, this local affine subspace approach can \commentTHa{represent the \acp{AIR} of} a larger class of acoustic scenes. However, its computational complexity \commentTHa{increases} significantly if the number of subspaces is increased to \commentTHc{represent} realistic acoustic scenes (cf.~Sec.~\ref{sec:air_mod_par_est}). Besides the affine subspace-based approaches, a globally nonlinear manifold model has been proposed in \cite{talmon_relative_2013} and \commentTHa{was} used in an offline least-squares system identification task. 
% \textcolor{red}{(write more why not taken into account)}.
% TODO: discuss why we did not use this one 
% In addition to the manifold-based approaches, prior knowledge about the \acp{AIR} has also been included by regularizing a least-squares cost function with the Mahalanobis distance based on an estimated \ac{AIR} covariance matrix \cite{fozunbal_multi-channel_2008}. As this approach does also assume a single global \ac{AIR} covariance matrix it exhibits the same limitations as the single affine subspace approach \commentTHa{(cf.~Sec.~\ref{sec:mix_mode})}.

In this paper, we introduce a novel algorithm which exploits an \commentTHd{adaptive \ac{AIR} subspace model} for enhancing \ac{KF}-based system identification algorithms. For this we \commentTHa{discuss} various \commentTHa{data-driven} local manifold approximations by affine subspaces.  
%and suggest a novel adaptive extension. 
The validity of the model assumptions is \commentTHd{verified} for simulated \commentTHa{\acp{AIR}}. 
% The \ac{AIR} subspace model is subsequently fused with a \ac{KF} for noise-robust \ac{OSASI}.
For fusing the \commentTHb{adaptive} subspace model with the \ac{KF} we suggest a novel probabilistic \commentTHa{inference} method to estimate the subspace parameters. Furthermore, we relax the idea of \commentTHe{hard projecting} a noisy \ac{AIR} estimate onto the \ac{AIR} manifold \commentTHd{to a soft projection} \commentTHe{which allows the algorithm to} cope with model imperfections. It is shown that the proposed \commentTHe{method} improves the convergence speed of \ac{KF}-based system identification \commentTHd{algorithms} and achieves higher steady-state performance in scenarios suffering from high-level interfering noise. \commentTHc{In addition, we show improved \ac{AF} performance for system identification scenarios that are challenged by nonunique \ac{MMSE} solutions \cite{sondhi_stereophonic_1995, sondhi_benesty_a_better_understanding}. This problem is often faced in rendering and teleconferencing applications for which the excitation signals are composed of less sources than loudspeakers.}

%\vspace*{5cm}
%in this paper we analyze the modelling capability of various affine subspace models  and we suggest
%online estimation of subspace model
%probabilistically motivated denoising of the filter estimates by exploiting Kalman filter estimation and the affine subspace estimate
%superior convergence properties for low SNR
%improved system identification performance for low-rank excitation
%We will show that this approach can also be used to deal with the nonuniquness problem (problem that algorithm does convergence against the true syste)

In this paper, \commentTHd{vectors are typeset} as bold \commentTHd{lowercase letters} and matrices as \commentTHc{bold uppercase} letters with underlined symbols representing time-domain quantities. The \commentTHd{all-zero} matrix of dimension $D_1 \times D_2$ is denoted by $\boldsymbol{0}_{D_1 \times D_2}$, the \commentTHa{$D \times D$}-dimensional identity matrix \commentTHa{by $\boldsymbol{I}_D$} and \commentTHa{the} \ac{DFT} matrix by $\boldsymbol{F}_D$, respectively. Transposition and Hermitian transposition are represented by $(\cdot)^{\text{T}}$ and $(\cdot)^{\text{H}}$. Furthermore, \commentTHa{the $i$th element of a vector $\boldsymbol{w}$ is denoted by $\left[\boldsymbol{w}\right]_i$ and} the $\text{diag}(\cdot)$ operator creates a diagonal matrix from its vector-valued argument. Finally, equivalency of two terms up to a constant is denoted by $\stackrel{\text{c}}{=}$.

The remainder of this paper is structured as follows: In Sec.~\ref{sec:probSigMod} a probabilistic signal observation model is introduced which relates the noisy observations to the unknown \acp{AIR}. Subsequently, in Sec.~\ref{sec:airAnalysis}, various affine subspace approaches to locally model an \ac{AIR} manifold are described and evaluated for simulated data. The fusion of the affine subspace models with a \ac{KF}-based \ac{OSASI} \commentTHe{algorithm} is introduced in Sec.~\ref{sec:kf}. Experimental results for the proposed algorithm are shown in Sec.~\ref{sec:experiments}. Finally, the paper is concluded in Sec.~\ref{sec:conclusion}.

\section{Probabilistic Signal Model}
\label{sec:probSigMod}
%\begin{itemize}
%	\item introduce here the probabilistic signal observation model
%	\item attention: only the state observation model and not the state transition model
%\end{itemize}

We \commentTHc{first introduce} a probabilistic signal model \commentTHc{for describing} the microphone observations \commentTHc{of a} \ac{MISO} system identification scenario with $B$ loudspeakers and one microphone as depicted in \makebox{Fig. \ref{fig:miso_sys_id}}. % \commentTHa{A straightforward} extension to multiple microphones is obtained by introducing several independent \ac{MISO} systems. 
\begin{figure}[tb] % TODO: maybe directly 
	\centering	
		\begin{tikzpicture}[node distance=1.5cm]
	\tikzset{loudspeaker style/.style={%
			draw,very thick,shape=loudspeaker,minimum size=5pt
	}}
	\tikzset{microphone style/.style={%
			draw,very thick,shape=microphone,minimum size=.5pt, inner sep=2.0pt
	}}
	
	\node (sigInX) at (0,.4) {};
% 	\node [left of=sigInX, node distance=2cm, inner sep=.0] (sigInX1) {};
	\node [right of=sigInX, inner sep=0] (spltX) {};
	\node [below of=spltX, rectangle, draw, thick, node distance=1.25cm] (pbkf) {AF};
	\node [below of=pbkf, circle, draw, inner sep=2, thick, node distance=1.9cm, minimum size=.35cm] (subtraction) {};
	\node (plusSymb) at (subtraction) {{\Large $+$}};
	% \node [right of=subtraction] (sigInY) {};
	% \node [rectangle, draw, thick] (pf)  at ($(subtraction)+(-4.0,.1)$) {PF};
	% \node [left of=subtraction, rectangle, draw, thick, node distance=3.25cm] (pf)  at ($(subtraction)+(-2,0)$) {PF};
	% \node [left of=pf, thick] (epf) {};
	% \node [left of=pbkf, rectangle, draw, node distance=2.2cm] (psd) {PSD Est.};
	\coordinate [left of=subtraction, inner sep=0, node distance=1.0cm]  (sigE) {};
	
	\node[loudspeaker style,rotate=-90] (speaker1) at (6.0,-.5) {};
	\node[loudspeaker style,rotate=-90] (speaker2) at (5.0,-.5) {};
%	\node (dots1) at (5.5, -.5) {$\dots$};
	\node (dots2) at (5.5, .0) {$\dots$};
	\node[microphone style,rotate=90] (mic) at (5.5,-2.25) {};
	
	\draw [thick] ($(sigInX)$) -| (speaker1.west) node [above, pos=.1] {{$\underline{\boldsymbol{x}}_{\tau}$}} node [pos=.76, right] {{$\underline{\boldsymbol{x}}_{B,\tau}$}};
	\draw [thick] (speaker2.west) -- ($(speaker2.west)+(0,.8)$) node [midway, left] {{$\underline{\boldsymbol{x}}_{1,\tau}$}};

	%\draw [->, thick] (sigInX) -| (pbkf);
	\draw [->, thick] (pbkf) -- (subtraction.north) node [pos=.4, right] {$\widehat{\underline{\boldsymbol{d}}}_{\tau}$} node [pos=.8, right] {$-$};
	\draw [->, thick] (mic.west) |- (subtraction.east) node [pos=.85, above] {$\underline{\boldsymbol{y}}_{\tau}$};
	\draw [->, thick] ($(pbkf.north)+(0,1.0)$) -- (pbkf.north);
	
	% \draw [-, thick] (subtraction.west) -- (sigE) node [pos=.3, above] {$\underline{\boldsymbol{e}}_{\tau}^+$};
	
	\draw [->, thick] (subtraction.west) -- (sigE) |- (pbkf.west) node [pos=.265, left] {$\underline{\boldsymbol{e}}_{\tau}^+$};

	% \draw [->, thick] (pf.west) -- (epf) node [pos=.5, above]  {$\hat{\underline{\boldsymbol{s}}}_{\tau}$};
	
	%\draw [-, thick] (sigInX1) -| (pbkf);
	% \draw [->, thick] ($(sigInX1)+(0.2,.0)$) |- ($(pf.east) + (0,.1)$);
	
	% \draw [->, thick] ($(pf.north)-(-0,0)$) |- (psd.west) node [pos=.3, left] {$\widehat{\boldsymbol{M}}_\tau$};
	% \draw [thick, <-] (psd.south) -- ($(psd.south)+(0,-1.65)$);
	% \draw [->, thick] (psd.east) -- (pbkf.west) node [pos=.5, above] {$\hat{\boldsymbol{\Psi}}_{\tau}^{(\cdot)}$};

	\draw [thick] (3.5,-2.55) rectangle (7,-.25);

	% \draw [thick] (subtraction.east) -| (mic.west);
	% \draw [thick] (sigInX1) -| (speaker1.west);
	
	%\draw [thick] ($(sigInX)+(-1.8,0)$) -- ($(sigInX)$);
	
	\draw [thick, dashed, ->] ($(speaker1)-(.0,.2)$) -- ($(mic)+(.0,.2)$);
	\draw [thick, dashed, ->] ($(speaker1)-(.0,.2)$) -- ($(speaker1)+(1.0,-.9)$) -- ($(mic)+(.1,.2)$);
	
	\draw [thick, dashed, ->] ($(speaker2)-(.0,.2)$) -- ($(mic)+(-.1,.2)$);
	
	% \node [] (d) at ($(mic)+(.5,0.7)$) {$\underline{\boldsymbol{d}}_\tau$}; 
	\node (d) at (6.1, -2.15) {${\underline{\boldsymbol{d}}}_{\tau}$};
	
	% \node [] (s) at ($(mic)+(-.55,.3)$) {$\boldsymbol{d}_\tau$};
	\draw [fill] (4.0,-1.4) circle (.05);
	\draw [thick] (3.97,-1.54) arc(265:375:.15);
	\node [] (s) at ($(4.60,-2.0)$) {$\underline{\boldsymbol{n}}_\tau$};
	\draw [dashed, ->, thick] ($(4.20,-1.53)+(0,0)$) -- ($(mic)+(-.15,.15)$);

%	\draw [fill] (3.8,-2.2) circle (.05);
%	\draw [thick] (3.9,-2.3) arc(-45:45:.15);
%	\draw [dashed, ->, thick] ($(4.0,-2.2)+(0,0)$) -- ($(mic)+(-.2,.10)$);
%	\node [] (s) at ($(4.5,-2.45)$) {$\underline{\boldsymbol{n}}_\tau$};
	
	\end{tikzpicture}
	\caption{Identification of an acoustic \ac{MISO} system.}	
	\label{fig:miso_sys_id}
\end{figure}

\commentTHe{A block of microphone observations at time index $\tau$}
\begin{equation}
	\underline{\boldsymbol{y}}_\tau = \underline{\boldsymbol{d}}_\tau + \underline{\boldsymbol{n}}_\tau \in \mathbb{R}^R
		\label{eq:td_sig_mod}
\end{equation}
\commentTHd{is modeled as a} superposition of the noise-free observation vector $\underline{\boldsymbol{d}}_\tau$ and the noise \commentTHd{signal} vector $\underline{\boldsymbol{n}}_\tau$. Each signal block, i.e., $\underline{\boldsymbol{y}}_\tau$, $\underline{\boldsymbol{d}}_\tau$, and $\underline{\boldsymbol{n}}_\tau$, contains $R$ consecutive samples:
\begin{equation}
	\underline{\boldsymbol{y}}_\tau = \begin{bmatrix}
	\underline{y}_{\tau R - R + 1}& \underline{y}_{\tau R - R + 2}& \dots& \underline{y}_{\tau R}
	\end{bmatrix}^{\trans} \in \mathbb{R}^{R}.
\end{equation}
%
%We assume a linear multi-path propagation model from the $P$ loudspeaker signals
%the noise-free observation is modelled by a linear convolutive FIr filter model
The noise-free observation $\underline{\boldsymbol{d}}_\tau$ is \commentTHa{modeled} by a linear convolution of the \acp{AIR} $\underline{\boldsymbol{w}}_{b,\tau} \in \mathbb{R}^{L}$ \commentTHd{with $b\in \{1,\dots,B\}$  of length $L$} with the respective loudspeaker signal blocks
\begin{equation}
\underline{\boldsymbol{x}}_{b,\tau} = \begin{bmatrix} \underline{{x}}_{b, \tau R-M+1} & \dots & \underline{{x}}_{b,\tau R}
\end{bmatrix}^{\trans} \in \mathbb{R}^M
\end{equation}
of length $M=R+L$ and subsequently adding up the $B$ convolution products. This can be \commentTHa{expressed} efficiently in the \ac{DFT} domain \cite{malik_recursive_2011}
\begin{equation}
	\underline{\boldsymbol{d}}_\tau = \sum_{b=1}^{B} \boldsymbol{Q}_1^{\trans} \boldsymbol{F}_M^{-1} \boldsymbol{X}_{b,\tau} \boldsymbol{F}_M \boldsymbol{Q}_2 \underline{\boldsymbol{w}}_{b,\tau},
	\label{eq:td_clean_sig_obs_eq}
\end{equation}
with the \commentTHa{diagonal matrix \makebox{$\boldsymbol{X}_{b,\tau} = \text{diag} \left( \boldsymbol{F}_M \underline{\boldsymbol{x}}_{b,\tau} \right)$}, containing the \commentTHc{\ac{DFT}-domain} loudspeaker signals of block $\tau$}, the constraint matrix \makebox{$\boldsymbol{Q}_1^{\trans} = \begin{bmatrix}\boldsymbol{0}_{R \times L} & \boldsymbol{I}_R\end{bmatrix}$} and the zero-padding matrix \commentTHb{\makebox{$\boldsymbol{Q}_2^{\trans}= \begin{bmatrix}\boldsymbol{I}_{L} & \boldsymbol{0}_{L \times R }\end{bmatrix}$}}. By inserting the convolution model \eqref{eq:td_clean_sig_obs_eq} into the additive signal model \eqref{eq:td_sig_mod} and introducing the \commentTHa{\acp{ATF}} \makebox{${\boldsymbol{w}}_{b,\tau} = \boldsymbol{F}_M \boldsymbol{Q}_2 \underline{\boldsymbol{w}}_{b,\tau}$}, we obtain the time-domain observation model
\begin{equation}
	\underline{\boldsymbol{y}}_\tau = \sum_{b=1}^{B} \boldsymbol{Q}_1^{\trans} \boldsymbol{F}_M^{-1} \boldsymbol{X}_{b,\tau} {\boldsymbol{w}}_{b,\tau} + \underline{\boldsymbol{n}}_\tau.
		%\underline{\boldsymbol{y}}_\tau = \sum_{b=1}^{B} \boldsymbol{Q}_1^{\trans} \boldsymbol{F}_M^{-1} \boldsymbol{X}_{b,\tau} \boldsymbol{F}_M \boldsymbol{Q}_2 \underline{\boldsymbol{w}}_{b,\tau} + \underline{\boldsymbol{n}}_\tau.
	\label{eq:td_obs_model}
\end{equation}
%
% a relationship between the noisy observations $\underline{\boldsymbol{y}}_\tau$ an the unknown \ac{FIR} filters $\underline{\boldsymbol{w}}_{\tau,p}$ with $p=1,\dots,P$. 
A corresponding frequency-domain observation model is computed by first \commentTHc{zero-padding} the signals in Eq.~\eqref{eq:td_obs_model} and subsequently applying \commentTHd{the} \ac{DFT} \cite{malik_recursive_2011}:
\begin{equation}
	{\boldsymbol{y}}_\tau =  \boldsymbol{F}_M \boldsymbol{Q}_1 {\underline{\boldsymbol{y}}}_{\tau} =  \boldsymbol{C}_{\tau}  {\boldsymbol{w}}_{\tau}  + {\boldsymbol{n}}_\tau \in \mathbb{C}^M.
	% {\boldsymbol{y}}_\tau = \sum_{b=1}^{B} \boldsymbol{C}_{b,\tau}  {\boldsymbol{w}}_{b,\tau}  + {\boldsymbol{n}}_\tau \in \mathbb{C}^M.
	% 	{\boldsymbol{y}}_\tau = \sum_{p=1}^{P} \boldsymbol{F}_M \boldsymbol{Q}_1 \boldsymbol{Q}_1^{\trans} \boldsymbol{F}_M^{-1} \boldsymbol{X}_{\tau,p}  {\boldsymbol{w}}_{\tau,p}  + {\boldsymbol{n}}_\tau
	\label{eq:fd_obs_eq}
\end{equation}
Here, we used the overlap-save constrained loudspeaker signal matrix 
\begin{equation}
\boldsymbol{C}_{\tau} = \begin{bmatrix}
\boldsymbol{C}_{1,\tau} & \dots & \boldsymbol{C}_{B,\tau}
\end{bmatrix} \in \mathbb{C}^{M \times BM}
\label{eq:broad_band_atf_def}
\end{equation}
with \makebox{$\boldsymbol{C}_{b,\tau}=\boldsymbol{F}_M \boldsymbol{Q}_1 \boldsymbol{Q}_1^{\trans} \boldsymbol{F}_M^{-1} \boldsymbol{X}_{b,\tau} \in \mathbb{C}^{M \times M}$} and the \ac{MISO} \commentTHa{\ac{ATF}} vector
\begin{equation}
{\boldsymbol{w}}_{\tau} = \begin{bmatrix} {\boldsymbol{w}}_{1,\tau}^{\trans} & \dots & {\boldsymbol{w}}_{B,\tau}^{\trans}
\end{bmatrix}^{\trans} \in \mathbb{C}^{MB}.
\label{eq:broad_band_fir_def}
\end{equation}
%
%with \makebox{${\boldsymbol{w}}_{b,\tau} = \boldsymbol{F}_M \boldsymbol{Q}_2 \underline{\boldsymbol{w}}_{b,\tau} \in \mathbb{C}^M$}.
Note that the corresponding time-domain \ac{AIR} vector is obtained by 
\begin{equation}
	\underline{\boldsymbol{w}}_\tau = \left( \boldsymbol{I}_B \otimes \left( \boldsymbol{Q}_2^{\trans} \boldsymbol{F}_M^{-1} \right) \right)\boldsymbol{w}_\tau \in \mathbb{R}^{Q} \label{eq:td_fir_filt_vec}
\end{equation}
with $Q=LB$.
%
%
%\begin{equation}
%	\boldsymbol{C}_{\tau} = \begin{bmatrix}
%	\boldsymbol{C}_{1,\tau} & \dots & \boldsymbol{C}_{B,\tau}
%	\end{bmatrix}
%\end{equation}
%
%\makebox{$\boldsymbol{C}_{b,\tau}=\boldsymbol{F}_M \boldsymbol{Q}_1 \boldsymbol{Q}_1^{\trans} \boldsymbol{F}_M^{-1} \boldsymbol{X}_{b,\tau}$} and the \ac{DFT}-transformed \ac{FIR} filters \makebox{${\boldsymbol{w}}_{b,\tau} = \boldsymbol{F}_M \boldsymbol{Q}_2 \underline{\boldsymbol{w}}_{b,\tau} \in \mathbb{C}^M$}. 
The interfering noise term ${\boldsymbol{n}_\tau}$ is modeled as a proper complex zero-mean Gaussian random vector \cite{malik_recursive_2011}
\begin{equation}
		 {\boldsymbol{n}_\tau} \sim   \mathcal{N}_c(\boldsymbol{n}_{\tau}|{\boldsymbol{0}_{M \times 1}}, \boldsymbol{\Psi}_{\tau}^{\text{N}}), \label{eq:obsMod}
\end{equation}
with covariance matrix $\boldsymbol{\Psi}_{\tau}^{\text{N}} \in \mathbb{C}^{M \times M}$.
% \textcolor{red}{Cite respective sources.}

% The time-domain observation model is straightforwardly transformed to the \ac{DFT} domain by first zero padding the signals

%By transforming the zero
%
%By inserting the convolution model Eq.~\eqref{eq:	\label{eq:td_clean_sig_obs_eq}} in the time-domain observation model \eqref{eq:td_obs_eq} and subsequently 

% \section{Acoustic Impulse Response Analysis}
\section{Analysis of Acoustic Impulse Responses}
 \label{sec:airAnalysis}
In the following sections, we discuss various approaches to model \commentTHd{the neighbourhood of the} unknown \ac{AIR} vector $\underline{\boldsymbol{w}}_\tau \in \mathbb{R}^Q$. We start by introducing the \commentTHc{first-order} Markov model assumption which is commonly used in \ac{KF}-based system identification algorithms and discuss its limitations. Subsequently, we describe how these limitations can be mitigated by modeling an \ac{AIR} manifold. Finally, we discuss various affine subspace-based approaches to locally approximate the manifold. 
%\commentTHa{Note that a straightforward extension of the respective \ac{MISO} models to a }
% Note that all \ac{MISO} \ac{AIR} models can be straightforwardly extended to \ac{MIMO} systems by stacking the respective \ac{MISO} \ac{AIR} vectors $\underline{\boldsymbol{w}}_\tau$ to an extended vector \cite{9231543}. 
\commentTHa{Note that a straightforward extension of the subsequently described \ac{MISO} \ac{AIR} models to \ac{MIMO} systems is obtained by stacking the respective \ac{MISO} \ac{AIR} vectors $\underline{\boldsymbol{w}}_\tau$ to an extended vector \cite{9231543}.}

%which is based on the assumption that only a structured subset of the high-dimensional space $\mathbb{R}^Q$ is populated.
% \textcolor{red}{maybe explain that a MIMO system can easily be exteneded by stacing multiple MISO systems}
%
%In the following we discuss various approaches to model the unknown \ac{FIR} filter vectors $\underline{\boldsymbol{w}}_\tau \in \mathbb{R}^Q$. They are based on the assumption that not all elements the \ac{FIR} vector space $\mathbb{R}^Q$ (cf.~Eq.~\eqref{eq:td_fir_filt_vec}) are equally likely, i.e., certain regions exhibit a higher probability. The extreme case that only a structured subset of the vector space $\mathbb{R}^Q$ is populated, motivates the assumption of a low-dimensional manifold as visualized in Fig.~\ref{fig:manifold_example}. % Its existence is often tightly coupled to the parameter changes of an underlying physical process, e.g., location or temperature, which governs the variability of the \ac{FIR} filter vectors. 
%
%\textcolor{red}{?????????? Or still keep Markov model here???????????? => NO}
%
%\vspace*{1cm}
%\begin{itemize}
%	\item \textcolor{red}{Affine subspace model and describe model with $V_i$ and $\bar{w}_i$} and describe projection
%	\item \textcolor{red}{Describe estimation of global model}
%	\item \textcolor{red}{Describe estimation of mixture model}
%	\item \textcolor{red}{Describe estimation of KNN model}
%\end{itemize}

\subsection{Acoustic Impulse Response Manifold}
\label{sec:markov_model}
In \cite{enzner_frequency-domain_2006, malik_recursive_2011} it is suggested \commentTHc{to} describe the temporal propagation of the \ac{DFT}-\commentTHc{domain} \ac{AIR} vector ${\boldsymbol{w}}_\tau$ (cf.~Eq.~\eqref{eq:td_fir_filt_vec}) by a random walk Markov model 
\begin{equation}
\begin{aligned}
{\w}_{\tau} &= \commentTHd{A}  ~{\w}_{\tau - 1}  + \Delta {{\w}}_{\tau}, \\
\Delta {{\w}}_{\tau} &\sim   \mathcal{N}_c(\Delta {{\w}}_{\tau}|{\boldsymbol{0}_{M \times 1}}, \boldsymbol{\Psi}_{\tau}^{\Delta \text{W}})
\end{aligned}
\label{eq:stateTransMod} 
\end{equation}
with the \commentTHa{state transition coefficient $ \commentTHd{A}$ and the} \commentTHa{block-diagonal} process noise covariance matrix
\begin{equation}
\boldsymbol{\Psi}_{\tau}^{\Delta \text{W}} = \begin{bmatrix} {\boldsymbol{\Psi}}_{11, \tau}^{\Delta \text{W}} & \dots & \boldsymbol{0}_{M \times M}\\
\vdots & \ddots & \vdots \\
\boldsymbol{0}_{M \times M}& \dots & {\boldsymbol{\Psi}}_{BB, \tau}^{\Delta W} 
\end{bmatrix}\in \mathbb{C}^{BM \times BM}. \label{eq:process_noise_cov_mat}
\end{equation}
As depicted in Fig.~\ref{fig:markov_model}, the random walk model describes a \commentTHb{continuous} temporal propagation of \commentTHa{subsequent \acp{AIR} $\underline{\boldsymbol{w}}_{\tau -1}$ and $\underline{\boldsymbol{w}}_{\tau}$, depicted as blue \commentTHd{dots}, of the \ac{OSASI} application.}
\begin{figure}[tb]
	\centering
	\begin{tikzpicture}[]
\pgfplotsset{
	colormap={blackwhite}{gray(0cm)=(0); gray(1cm)=(.9)}
}

\filldraw[even odd rule,inner color=gray,outer color=white, dashed] (1.3,1.3,1)  circle (1.2);
%\draw(0,0) circle (1.8);
\draw [dashed] (0.1,1.3,1) arc (180:540:1.2 and 0.6);

\coordinate (center) at (0.0, 0, 0.0);

\draw [thick, ->] ($(center)+(0,0,0)$) -- ($(center)+(2,0,0)$) node [at end, right] {$[\underline{\boldsymbol{w}}_{(\cdot)}]_2$};
\draw [thick,->] ($(center)+(0,0,0)$) -- ($(center)+(0,2,0)$) node [at end, left] {$[\underline{\boldsymbol{w}}_{(\cdot)}]_3$};
\draw [thick,->] ($(center)+(0,0,0)$) -- ($(center)+(0,0,2)$) node [at end, left] {$[\underline{\boldsymbol{w}}_{(\cdot)}]_1$};

%	\draw [circle, fill,color=black] (1.3,1.3,1) circle (2pt);
%	\node at (1.2, 1.2, 1.5) {\textcolor{black}{$\underline{\boldsymbol{w}}_{\tau-1}$}};
%	
%	\draw [circle, fill,color=black] (1.8,2.1,1) circle (2pt);
%	\node at (1.4, 2.15, 1.0) {\textcolor{black}{$\underline{\boldsymbol{w}}_{\tau}$}};
	
	\draw [circle, fill,color=blue] (1.3,1.3,1) circle (2pt);
	\node at (2.1, 1.45, 1.5) {\textcolor{black}{$\underline{\boldsymbol{w}}_{\tau-1}$}};
	
	\draw [circle, fill,color=blue] (1.8,2.1,1) circle (2pt);
	\node at (1.4, 2.15, 1.0) {\textcolor{black}{$\underline{\boldsymbol{w}}_{\tau}$}};
	
	\draw [circle, fill,color=black] (0.55,0.72,1) circle (2pt);
	\draw [circle, fill,color=black] (0.75,0.85,1) circle (2pt);
	\draw [circle, fill,color=black] (0.95,0.95,1) circle (2pt);
	\draw [circle, fill,color=black] (1.15,1.1,1) circle (2pt);
	
	\draw [circle, fill,color=black] (1.45,1.5,1) circle (2pt);
	\draw [circle, fill,color=black] (1.6,1.7,1) circle (2pt);
	\draw [circle, fill,color=black] (1.725,1.87,1) circle (2pt);
	
	\draw [circle, fill,color=black] (1.88,2.33,1) circle (2pt);
	\draw [circle, fill,color=black] (1.95,2.6,1) circle (2pt);

%	\begin{scope}[xshift=6cm]
%	\filldraw[even odd rule,inner color=red,outer color=red!5] (0,0) circle (1.8);
%	\draw(0,0) circle (2.2);
%	\end{scope}
	
  %\shade[ball color = gray!40, opacity = 0.4] (1.3, 1.3, 1) circle (1.2cm);
%	\draw (1.3, 1.3, 1) circle (1.2cm);
%	\draw [dashed] (1.3, 1.3, 1) circle (.8cm);
%	\draw [dashed] (1.3, 1.3, 1) circle (.4cm);

\end{tikzpicture}
	%\caption{\commentTHa{text}}
	\caption{Exemplary \acp{AIR} \commentTHd{of length $Q=3$} of an acoustic scene with blue dots representing subsequent \acp{AIR} of the \ac{OSASI} application. The shaded gray \commentTHd{ball, delimited with dashed contour lines, depicts an exemplary process noise covariance matrix around $\underline{\boldsymbol{w}}_{\tau-1}$.}}	
	%		Visualizing two subsequent \acp{AIR} $\underline{\boldsymbol{w}}_{\tau-1}$ and $\underline{\boldsymbol{w}}_\tau$ as blue dots including exemplary samples of 		
	%		. The shaded gray area depicts an exemplary process noise covariance matrix around the \ac{AIR} $\underline{\boldsymbol{w}}_{\tau-1}$ including its dashed contour lines. The black dots represent exemplary samples of surrounding \acp{AIR} which are confined to a one-dimensional \ac{AIR} manifold.}
	% \caption{\commentTHa{The random walk Markov model enforces a smooth temporal propagation of subsequent \acp{AIR}, e.g., $\underline{\boldsymbol{w}}_\tau$ and $\underline{\boldsymbol{w}}_\tau$, shown as blue dots ($Q=3$). The shaded gray area depicts an exemplary process noise covariance matrix around the \ac{AIR} $\underline{\boldsymbol{w}}_\tau$ including the dashed contour lines. The black dots represent exemplary samples of surrounding \ac{AIR} vectors which are confined to an \ac{AIR} manifold.}}
	%\caption{The Markov model enforces a smooth trajectory of estimates by a random walk state transition model. The shaded gray area depicts an exemplary process noise covariance matrix ($Q=3$).}
	\label{fig:markov_model}
\end{figure}
%
%  This model enforces a smooth  trajectory of subsequent \ac{AIR} and thus counteracts outliers (cf.~Fig.~\ref{fig:markov_model})
However, \commentTHe{as} no additional knowledge about the \commentTHb{filter coefficients} is assumed, the process noise \commentTHa{power} is distributed into all directions of the \commentTHa{high-dimensional} \commentTHe{\ac{FIR} filter vector space $\mathbb{R}^Q$,} \commentTHa{as shown by the shaded gray area in Fig.~\ref{fig:markov_model}}. In contrast, the \acp{AIR} in the \commentTHa{vicinity} of \commentTHb{$\underline{\boldsymbol{w}}_{\tau-1} \in \mathbb{R}^Q$} \commentTHe{often populate only a subset of this space} \cite{laufer-goldshtein_study_2015}. This is visualized in Fig.~\ref{fig:markov_model} by showing \commentTHa{exemplary samples of} the surrounding \ac{AIR} vectors as black dots.
% This motivates the idea of modelling each neighbourhood by an affine subspace
On a global view this motivates the assumption that all \acp{AIR} are confined to a structured subset of the vector space $\mathbb{R}^Q$ which is termed \ac{AIR} manifold. As manifolds are locally Euclidean \cite{tu2010introduction}, each neighbourhood \commentTHb{of an \ac{AIR}} can be described by an affine subspace $\commentTHe{\mathcal{M}_i}$ of the vector space $\mathbb{R}^Q$. This is visualized in Fig.~\ref{fig:manifold_example} where each \commentTHa{shaded grid cell} \commentTHa{illustrates} a different affine subspace $\mathcal{M}_i$. \commentTHe{We now} introduce a mathematical description of a single affine subspace which will serve as a basis for the following approaches to describe the manifold globally by patches of affine subspaces.  % In the following we describe various affine subspace-based approaches to model the manifold globally. 

% In particular during the convergence phase and in noisy applications it would be desirable to distribute the uncertainty in the direction of likely estimates. \textcolor{red}{(Maybe move this to another section!)}
% In particular during the convergence phase and in noisy applications it would be desirable to distribute the uncertainty in the direction of likely estimates. \textcolor{red}{(Maybe move this to another section!)}

%However, In particular during the convergence phase and in noisy applications it would be desirable to distribute the uncertainty in the direction of the optimum estimate.
%
%
%This limits the convergence and steady-state performance in noisy applications 
%This is problematic during the convergence phase and in noisy applications as highly unlikely
%Thus, under the assumption that not all not all elements of the $Q$-dimensional time domain \ac{FIR} filter space (cf.~Eq.~\eqref{eq:td_fir_filt_vec}) are equally likely, i.e., certain regions exhibit a higher probability, 
%Thus, even highly unlikely areas
%\textcolor{red}{Describe limitations!}
%

\subsection{Affine Subspace Model}
\label{sec:aff_subspace_models}
An affine subspace $\mathcal{M}_i$ of dimension $D_i$ is defined by:
\begin{equation}
	\mathcal{M}_i := \{ \underline{\boldsymbol{w}}_\tau^{\text{p}} \in \mathbb{R}^Q \vert \underline{\boldsymbol{w}}_\tau^{\text{p}}=\underline{\bar{\boldsymbol{\boldsymbol{w}}}}_{i} + \underline{{\boldsymbol{V}}}_{i} \underline{\boldsymbol{\beta}}_\tau,~ \underline{\boldsymbol{\beta}}_\tau \in \mathbb{R}^{D_i} \}.
	\label{eq:aff_subspace_def}
	% \mathcal{M}_i := \{ \boldsymbol{w}_\tau^{\text{p}}= \underline{\bar{\boldsymbol{\boldsymbol{w}}}}_{i} + \underline{{\boldsymbol{V}}}_{i} \boldsymbol{\beta}_i \vert~ \boldsymbol{\beta}_i \in \mathbb{R}^{D_i} \}
\end{equation}
It is parametrized by its offset $\bar{\underline{\boldsymbol{w}}}_i \in \mathbb{R}^Q$ and its basis matrix $\underline{\boldsymbol{V}}_i \in \mathbb{R}^{Q \times D_i}$. 
% Note that if $D_i=R$ and the columns of $\underline{\boldsymbol{V}}_i$ are a basis of the $\mathbb{R}^Q$, the coefficient vector $\beta_\tau$ represents just a non-canonical basis representation.
The vector \commentTHd{$\underline{\boldsymbol{\beta}}_\tau$} represents the coordinates of the affine subspace element $\underline{\boldsymbol{w}}_\tau^{\text{p}}$ in the basis spanned by the columns of $\underline{\boldsymbol{V}}_i$.

An orthogonal projection of an arbitrary \ac{AIR} vector $\underline{\boldsymbol{w}}_\tau \in \mathbb{R}^Q$ onto the affine subspace $\mathcal{M}_i$ is given by the linear mapping \cite{strang2006linear}
\begin{align}
\underline{\boldsymbol{w}}_\tau^{\text{p}}  = f_{\mathcal{M}_i} ({\underline{\boldsymbol{w}}}_\tau) 
 = \bar{\underline{\boldsymbol{w}}}_i + \commentTHb{\underline{\boldsymbol{L}}_i}  \left( {\underline{\boldsymbol{w}}}_\tau - \bar{\underline{\boldsymbol{w}}}_i \right) \in \mathbb{R}^{ Q}
\label{eq:projEstimate}
\end{align}
with the rank-$D_i$ projection matrix 
\begin{equation}
\commentTHb{\underline{\boldsymbol{L}}_i} = \underline{\boldsymbol{V}}_i \left(\underline{\boldsymbol{V}}_i^{\text{T}} \underline{\boldsymbol{V}}_i\right)^{-1} \underline{\boldsymbol{V}}_i^{\text{T}} \in \mathbb{R}^{Q \times Q}.
\label{eq:defProj}
\end{equation}
\commentTHe{Here, it is assumed that the columns in $\underline{\boldsymbol{V}}_i$ are linearly independent.}
Fig.~\ref{fig:manifold_example} shows the projection of the \ac{AIR} vector $\underline{\boldsymbol{w}}_\tau$ onto an exemplary \ac{AIR} manifold which is locally represented by the affine subspace $\mathcal{M}_i$. % In the following section we will discuss how to approximate the manifold globally using the affine subspace definition \eqref{eq:aff_subspace_def} and how to estimate its parameters from a set of $G$ training samples $\underline{\boldsymbol{w}}_\kappa^{\text{tr}}$ with $g=1,\dots,G$. 
%
%estimate the parameters, i.e., the offset $\bar{\underline{\boldsymbol{w}}}_i$ and the basis matrix $\underline{\boldsymbol{V}}_i $, of the affine subspace.
%
\begin{figure}[tb]
	%\documentclass[tikz,border=3.14mm]{standalone}
%\usepackage{pgfplots}
%\pgfplotsset{compat=1.15}
%\usetikzlibrary{calc}
%\begin{document}

\begin{tikzpicture}[]
\pgfplotsset{
	colormap={blackwhite}{gray(0cm)=(.4); gray(1cm)=(.9)}
}

\begin{axis}[
hide axis,
%view/h=45,
%xlabel = {$x$},
%ylabel = {$y$},
%zlabel = {$z$},
declare function={f(\x,\y)=10-(\x^2+\y^2);},		%f(x,y)
declare function={c_x(\t)=cos(\t)+1;},				% cx(t)
declare function={c_y(\t)=sin(\t)-1;},				% cy(t)
declare function={c_z(\t)=f(c_x(\t),c_y(\t));},zmax=9.25, samples=12]	% cz(t) = f(cx(t), cy(t))
\addplot3[surf,  domain=-1.7:1.7,domain y=-2:2,]{f(x,y)};
%\addplot3[black,opacity=1.0,variable=t,domain=0:360] ({c_x(t)},{c_y(t)},{c_z(t)});

\coordinate (center) at (150.0, 40, -0.0);

\draw [thick, ->] ($(center)+(0,0,0)$) -- ($(center)+(60,0,0)$) node [at end, right] {$[\underline{\boldsymbol{w}}_{(\cdot)}]_2$};
\draw [thick,->] ($(center)+(0,0,0)$) -- ($(center)+(0,-98,0)$); % node [at end, left,above] {$[\underline{\boldsymbol{w}}]_1$};
\draw [thick,->] ($(center)+(0,0,0)$) -- ($(center)+(0,0,160)$) node [at end, left] {$[\underline{\boldsymbol{w}}_{(\cdot)}]_3$};

\node (w1) at ($(center)+(-35,-70,20)$) {$[\underline{\boldsymbol{w}}_{(\cdot)}]_1$};

	%\draw[fill=white,fill opacity=0.8, tangent plane=at {(-.75,-.75,10)} with vectors {(1.5,0,0)} and {(0,1.5,0)}];
	% \draw[fill=white,fill opacity=0.8, tangent plane=at {(-1.25,-1.0,5)} with vectors {(1.5,0, 0)} and {(0,1,2.5)}];

%\draw[fill=white,fill opacity=0.8,tangent plane=at {(.5,-1.0,5)} with vectors {(1,2,-1.5)} and {(-1,1,2)}];

% \node at (-1.0,-.8, 5.4) {$\mathcal{M}_1$};

%	\node at (.0,0,10) {$\mathcal{M}_2$};

%	\draw [->, thick, red] (-0.4,-.8, 6.5) -- (0.3,-.8, 6.5);
%	\draw [->, thick, red] (-0.4,-.8, 6.5) -- (-.4,-.4, 7.5);

\end{axis}

	\coordinate (est) at (2.9, 5.5, 0.2);
\draw [circle, fill, red] ($(est)+(0,0,0)$) circle (2pt);
\node at ($(est)+(.4,.0,0)$) {\textcolor{black}{$\underline{\boldsymbol{w}}_{\tau}$}};

	\coordinate (proj) at (2.9, 4.0, 0.2);
	
\draw [dashed, thick, ->] ($(est)+(0,-.1, 0)$) -- (proj);

	\draw [circle, fill,blue] (2.0, 3.2, 0.0) circle (2pt);
	\draw [circle, fill,blue] (2.9, 3.9, 0.2) circle (2pt);
	\draw [circle, fill,blue] (3.9, 4.2, 0.2) circle (2pt);

	\node at (2.6, 3.1, 0.0) {\textcolor{black}{$\underline{\boldsymbol{w}}_{\tau-1}^{\text{p}}$}};
	\node at (3.3, 3.8, 0.2) {\textcolor{black}{$\underline{\boldsymbol{w}}_{\tau}^{\text{p}}$}};
	\node at (4.5, 4.1, 0.2) {\textcolor{black}{$\underline{\boldsymbol{w}}_{\tau+1}^{\text{p}}$}};
	
		\node at (1.5,4.8, 0) {$\mathcal{M}_i$};
		\draw [dashed, thick] ($(1.3,4.6, 0)+(.2,0,0)$) to [out=-60,in=-180] ($(proj)+(-.1,-.3,0)$);

	\node at (6.1,1.8, 0) {$\mathcal{M}_2$};
	\draw [dashed, thick] ($ (6.0,1.6, 0) +(-1.2,0,0)$) to [out=-0,in=-180]  ($ (6.1,1.8, 0) +(-.35,-.0,0)$);
		
	\node at (6.1,1.1, 0) {$\mathcal{M}_1$};
\draw [dashed, thick] ($ (6.0,0.8, 0) +(-1.4,0,0)$) to [out=-0,in=-180]  ($ (6.1,1.1, 0) +(-.35,-.0,0)$);
		
	\node at (6.05,2.4, 0) {$\boldsymbol{\vdots}$};	
\end{tikzpicture}

%\end{document}
	\caption{Projection of an \ac{AIR} $\underline{\boldsymbol{w}}_\tau$ \commentTHd{of length $Q=3$} onto an exemplary \ac{AIR} manifold. Each \commentTHe{shaded} grid cell represents an affine subspace $\mathcal{M}_i$ \commentTHc{of dimension $D_i=2$} which is locally tangential to the manifold.} %Note that in contrast to Fig.~\ref{fig:markov_model} \commentTHa{a two-dimensional \ac{AIR} manifold is visualized.}}
	\label{fig:manifold_example}
\end{figure}

\subsection{Affine Subspace Parameter Estimation}
\label{sec:par_est}
We now discuss how to estimate the parameters of a single affine subspace $\mathcal{M}_i$, i.e., its offset $\bar{\underline{\boldsymbol{w}}}_{i}$ and basis matrix $\underline{\boldsymbol{V}}_{i}$, from a local training data set $\mathcal{U}_i$ including \commentTHe{$K_i$} \ac{AIR} vectors \commentTHc{$\underline{\boldsymbol{w}}_\kappa^{\text{tr}}$}. We assume that the \commentTHe{$K_i$} samples of the local training data set $\mathcal{U}_i$ \commentTHd{form} a subset of the full training data set $\mathcal{U}$\commentTHb{, i.e., $\mathcal{U}_i  \subseteq  \mathcal{U}$,} consisting of \commentTHe{$K$} samples in total. The indicator variable 
\begin{equation}
\commentTHc{z_{\kappa i}} :=
\begin{cases}
1 &\text{if } \underline{\boldsymbol{w}}_\kappa^{\text{tr}} \in \mathcal{U}_i, \\
0 &\text{if } \underline{\boldsymbol{w}}_\kappa^{\text{tr}} \notin \mathcal{U}_i
\end{cases}
\end{equation}
describes the assignment of the \commentTHa{training} samples to the local training data subsets. \commentTHd{The} offset of the affine subspace $\mathcal{M}_i$ \commentTHd{is estimated} as arithmetic average of the local training data:
\begin{equation}
\commentTHc{\bar{\underline{\boldsymbol{w}}}_i = \frac{1}{\commentTHe{K_i}} \sum_{\kappa=1}^{\commentTHe{K}} z_{\kappa i} \underline{\boldsymbol{w}}_\kappa^{\text{tr}}}
\label{eq:mean_comp}
\end{equation}
with \commentTHc{$\commentTHe{K_i} = \sum_{\kappa=1}^{\commentTHe{K}} z_{\kappa i}$.}
%
% A very popular method for deducing the basis is given by \ac{PCA} framework. For this we first estimate the local \ac{AIR} covariance matrix
For computing the basis matrix $\underline{\boldsymbol{V}}_i$ we first estimate the local \ac{AIR} covariance matrix
%A very popular method for deducing a basis matrix $\underline{\boldsymbol{V}}_i$ is to first estimate the \ac{AIR} covariance matrix
\begin{equation}
\commentTHe{\underline{\boldsymbol{R}}_{i} = \frac{1}{K_i - 1} \sum_{\kappa=1}^{\commentTHe{K}} z_{\kappa i} (\underline{\boldsymbol{w}}_\kappa^{\text{tr}} - \bar{\underline{\boldsymbol{\boldsymbol{w}}}}_i) {(\underline{\boldsymbol{w}}_\kappa^{\text{tr}} - \bar{\underline{\boldsymbol{\boldsymbol{w}}}}_i)}^{\text{T}}. } %\in \mathbb{R}^{Q \times Q} % = \boldsymbol{U} \boldsymbol{D} \trans{\boldsymbol{U}}
%\boldsymbol{C}_i = \mathbb{E} \left[(\vecModMIMO_i - \bar{\boldsymbol{h}}_i) \trans{(\vecModMIMO_i - \bar{\boldsymbol{h}}_i)} \right].
\label{eq:test}
\end{equation}
Subsequently, the basis matrix $\underline{\boldsymbol{V}}_i$ is \commentTHa{determined} by the eigenvectors corresponding to the \commentTHc{$D_i$ largest} eigenvalues. Note that, due to the broadband \commentTHc{nature of the \ac{AIR} vector \eqref{eq:td_fir_filt_vec}, the covariance matrix $\underline{\boldsymbol{R}}_i$ describes the correlation between different \acp{AIR}, i.e., $\underline{\boldsymbol{w}}_{1,\tau},~\dots,~\underline{\boldsymbol{w}}_{B,\tau}$ \commentTHd{as well as} the correlation of different taps of one \ac{AIR} $\underline{\boldsymbol{w}}_{b,\tau}$.} We now discuss how the number of affine subspaces $I$ and the selection of the indicator variables \commentTHc{$z_{\kappa i}$} results in different approaches to approximate the \ac{AIR} manifold.

\subsection{Local Training Data Estimation}
\label{sec:air_mod_par_est}
% TODO: check that data sets are really non onverlapping
In \cite{9231543} it is proposed to cluster the training data into $I$ \commentTHa{disjoint} local data sets $\mathcal{U}_i$ by the k-means algorithm \cite{lloyd_least_1982, Arthur07}. Subsequently, each local training data set $\mathcal{U}_i$ is used to compute a \commentTHd{local affine subspace $\mathcal{M}_i$} by the method described in Sec.~\ref{sec:par_est}. \commentTHd{Note that the classical global \ac{PCA} approach is contained as a special case with $I=1$.} 
\commentTHc{We now} evaluate the validity of this model for a typical \ac{MISO} acoustic rendering scenario. The simulation parameters are summarized in Sec.~\ref{sec:experiments} and the models are learned from $\commentTHe{K}=5000$ training data samples. 
As evaluation measure we use the logarithmic system mismatch % \cite{enzner_acoustic_2014}
\begin{equation}
\Upsilon_\tau = 10\log_{10} \left( \frac{1}{B} \sum_{b=1}^{B} \frac{||{\underline{\boldsymbol{w}}_{b,\tau} - \boldsymbol{Q}_3 \underline{\hat{\boldsymbol{w}}}_{b,\tau}^{\text{p}}||^2}}{||{\underline{\boldsymbol{w}}}_{b,\tau}||^2}\right)
\label{eq:systMisDef1}
\end{equation}
of the true \commentTHe{\acp{AIR}} $\underline{\boldsymbol{w}}_{b,\tau} \in \mathbb{R}^{W}$ of length \commentTHc{$W \geq L$} and the \commentTHe{projections} of its first $L$ taps $\underline{\hat{\boldsymbol{w}}}_{b,\tau}^{\text{p}} \in \mathbb{R}^L$ onto the \commentTHa{respective} affine subspace \commentTHa{model}. Note that the \commentTHc{zero-padding} matrix \makebox{$\boldsymbol{Q}_3^{\trans}= \begin{bmatrix}\boldsymbol{I}_{L} & \boldsymbol{0}_{L \times W-L }\end{bmatrix} \in \mathbb{R}^{L \times W}$} ensures \commentTHc{equal lengths} of the true \ac{AIR} \commentTHc{vector} and its \commentTHd{truncated} projection. Fig.~\ref{fig:subspace_system_mismatch} shows the average system mismatch \commentTHe{$\bar{\Upsilon}$} which results from \commentTHc{the projection of $500$ ground-truth \ac{AIR} vectors $\underline{\boldsymbol{w}}_\tau$ onto the global affine subspace model \makebox{(\textit{Global Proj.} ($I=1$))} and a mixture model with \makebox{$I=40$} clusters \makebox{(\textit{Mixture Proj.} ($I=40$))} in dependence of the subspace dimension $D_i$}. Note that for the mixture model the affine subspace with the lowest system mismatch $\Upsilon_\tau$ is selected.
% \commentTHa{which requires oracle knowledge about the true \ac{AIR}}. 
In addition a \commentTHc{benchmark, \textit{Oracle \commentTHc{GT}},} is evaluated which \commentTHd{represents the optimum choice for a length-$L$ \commentTHe{FIR filter} representation by using the first $L$ samples of the ground-truth \acp{AIR}}, i.e., $\underline{\hat{\boldsymbol{w}}}_{b,\tau}^{\text{p}} = \boldsymbol{Q}_3^T \underline{{\boldsymbol{w}}}_{b,\tau} $.
\begin{figure}[tb]
	\newlength\fwidth
	\setlength\fwidth{.7\columnwidth}
\definecolor{mycolor1}{rgb}{1.00000,0.7,0.00000}%
\definecolor{mycolor2}{rgb}{0.00000,0.34483,0.00000}%
% \definecolor{mycolor3}{rgb}{1.00000,0.10345,0.72414}%
\definecolor{mycolor4}{rgb}{0.00000,0.00000,0.17241}%
\definecolor{mycolor5}{rgb}{0.00000,0.4,0.2}%
\definecolor{myMagenta}{rgb}{0.7,0.0,0.9}%

\def\liwi{1pt}
\begin{tikzpicture}

\begin{axis}[%
width=0.951\fwidth,
height=0.725\fwidth,
at={(0\fwidth,0\fwidth)},
scale only axis,
xmin=0,
xmax=1000,
xlabel style={font=\color{white!15!black}},
xlabel={Subspace dimension $D_i$},
ymin=-18,
ymax=0,
% extra x ticks = {120},
ylabel style={font=\color{white!15!black}},
ylabel={\commentTHe{$\bar{\Upsilon}$} in dB},
axis background/.style={fill=white},
axis x line*=bottom,
axis y line*=left,
xmajorgrids,
ymajorgrids,
legend style={legend cell align=left, align=left, draw=white!15!black, at={(0.42,1.3)},anchor=north, legend columns=2}
]
\addplot [color=myMagenta, line width=\liwi]
  table[row sep=crcr]{%
1	-17.05\\
1024	-17.05\\
};
\addlegendentry{Oracle \commentTHc{GT}}

\addplot [color=mycolor1, line width=\liwi]
  table[row sep=crcr]{%
1	-0.9934\\
2	-1.417\\
3	-1.832\\
5	-2.631\\
6	-2.901\\
7	-3.153\\
8	-3.391\\
9	-3.518\\
10	-3.637\\
20	-4.38\\
30	-4.88\\
40	-5.297\\
50	-5.703\\
60	-6.071\\
70	-6.428\\
80	-6.753\\
90	-7.059\\
100	-7.347\\
110	-7.626\\
130	-8.145\\
140	-8.392\\
160	-8.861\\
180	-9.295\\
190	-9.51\\
200	-9.716\\
210	-9.93\\
220	-10.13\\
230	-10.33\\
240	-10.53\\
280	-11.3\\
290	-11.48\\
310	-11.84\\
320	-12.02\\
380	-13.04\\
430	-13.83\\
460	-14.27\\
480	-14.56\\
490	-14.7\\
530	-15.24\\
580	-15.83\\
600	-16.04\\
620	-16.24\\
640	-16.43\\
660	-16.6\\
670	-16.68\\
690	-16.81\\
700	-16.86\\
720	-16.94\\
730	-16.97\\
750	-17.01\\
770	-17.03\\
800	-17.04\\
920	-17.05\\
1020	-17.05\\
};
\addlegendentry{Global \commentTHc{Proj.} ($I=1$)}

\addplot [color=red, line width=\liwi]
table[row sep=crcr]{%
	0	-6.744\\
	1	-8.237\\
	2	-8.981\\
	3	-9.461\\
	4	-9.88\\
	6	-10.41\\
	7	-10.6\\
	8	-10.76\\
	9	-10.92\\
	19	-11.86\\
	29	-12.38\\
	39	-12.74\\
	49	-13.02\\
	59	-13.25\\
	69	-13.46\\
	79	-13.65\\
	89	-13.82\\
	99	-13.98\\
	119	-14.28\\
	139	-14.54\\
	149	-14.67\\
	169	-14.88\\
	179	-14.98\\
	189	-15.09\\
	199	-15.19\\
	219	-15.37\\
	239	-15.53\\
	269	-15.74\\
	289	-15.88\\
	319	-16.05\\
	359	-16.26\\
	379	-16.34\\
	419	-16.5\\
	449	-16.61\\
	469	-16.67\\
	509	-16.78\\
	559	-16.89\\
	599	-16.95\\
	639	-16.99\\
	679	-17.02\\
	729	-17.04\\
	789	-17.05\\
	1019	-17.05\\
};
\addlegendentry{KNN \commentTHc{Proj.}}

\addplot [color=mycolor4, line width=\liwi]
  table[row sep=crcr]{%
1	-3.597\\
2	-3.885\\
3	-4.107\\
4	-4.305\\
5	-4.47\\
7	-4.749\\
8	-4.881\\
9	-4.994\\
10	-5.124\\
20	-6.229\\
30	-7.204\\
40	-8.046\\
50	-8.773\\
60	-9.404\\
70	-10\\
80	-10.53\\
90	-11.03\\
100	-11.46\\
110	-11.83\\
120	-12.15\\
130	-12.34\\
140	-12.45\\
150	-12.55\\
160	-12.61\\
170	-12.66\\
190	-12.75\\
220	-12.88\\
250	-13.03\\
330	-13.44\\
400	-13.82\\
590	-14.86\\
730	-15.65\\
770	-15.88\\
830	-16.22\\
900	-16.6\\
940	-16.79\\
970	-16.92\\
990	-17\\
1000	-17.03\\
};
\addlegendentry{Mixture \commentTHc{Proj.} ($I=40$)}

%\addplot [color=green, line width=\liwi]
%table[row sep=crcr]{%
%	1	-6.744\\
%	1024	-6.744\\
%};
%\addlegendentry{NN Projection}

%\addplot [color=red, line width=\liwi]
%table[row sep=crcr]{%
%	0	-6.744\\
%	1	-8.237\\
%	2	-8.981\\
%	3	-9.461\\
%	4	-9.88\\
%	6	-10.41\\
%	7	-10.6\\
%	8	-10.76\\
%	9	-10.92\\
%	19	-11.86\\
%	29	-12.38\\
%	39	-12.74\\
%	49	-13.02\\
%	59	-13.25\\
%	69	-13.46\\
%	79	-13.65\\
%	89	-13.82\\
%	99	-13.98\\
%	119	-14.28\\
%	139	-14.54\\
%	149	-14.67\\
%	169	-14.88\\
%	179	-14.98\\
%	189	-15.09\\
%	199	-15.19\\
%	219	-15.37\\
%	239	-15.53\\
%	269	-15.74\\
%	289	-15.88\\
%	319	-16.05\\
%	359	-16.26\\
%	379	-16.34\\
%	419	-16.5\\
%	449	-16.61\\
%	469	-16.67\\
%	509	-16.78\\
%	559	-16.89\\
%	599	-16.95\\
%	639	-16.99\\
%	679	-17.02\\
%	729	-17.04\\
%	789	-17.05\\
%	1019	-17.05\\
%};
%\addlegendentry{KNN \commentTHc{Proj.}}

\end{axis}
\end{tikzpicture}%
	\caption{\commentTHd{Average system mismatch \commentTHe{$\bar{\Upsilon}$} of projecting $500$ ground-truth \ac{MISO} \ac{AIR} vectors $\underline{{\boldsymbol{w}}}_{\tau} \in \mathbb{R}^{BW}$ onto various affine subspace models ($B=2$, $W=4096$, $L=512$). The subspace dimension $D_i=0$ corresponds to using only the offset vector of the model as projected \ac{AIR}.}}
	\label{fig:subspace_system_mismatch}
\end{figure}
We conclude from Fig.~\ref{fig:subspace_system_mismatch} that for the considered scenario the global affine subspace assumption holds only \commentTHa{coarsely}. Due to the high variability \commentTHc{of the \acp{AIR} representing the acoustic scene\commentTHd{,} a large subspace dimension} is required to attain a reasonable average  system mismatch \commentTHe{$\bar{\Upsilon}$}. This limits the denoising capability of the respective projection \commentTHe{(cf.~Sec.~\ref{sec:softSubProj})}. In contrast\commentTHd{,} the mixture approach attains \commentTHc{for small subspace dimensions $D_i$ already} a much lower average system mismatch \commentTHe{$\bar{\Upsilon}$}. However, as the affine subspaces $\mathcal{M}_i$ are only representative for samples \commentTHd{close to the offset vector} $\bar{\underline{\boldsymbol{w}}}_i$, its modeling capability depends decisively on the number of clusters $I$. % \textcolor{red}{We observe that for $I=40$ clusters the best attainable system mismatch is limited to approximately $\bar{\Upsilon}_\tau=x$ dB for the considered scenario. Sharp knick at $120$ which is due to the non overlapping assumption }
%
%
%This is due to the coarse modelling  
%
%
%For each experiment the cluster with the lowest projected system mismatch is selected. We conclude from \ref{fig:subspace_system_mismatch} that the local approach attains at an equal subspace dimension a much lower system mismatch. However, the local models are only representative for samples close the local mean. Thus, the modelling capability of the mixture approach highly depends on the number of clusters. We observe that for $I=40$ clusters the best attainable system mismatch is limited to approximately $\bar{\Upsilon}=x$ dB for the considered scenario. 
%
% \subsubsection{KNN-based Models}
% \label{sec:knn_aff_subspace}
This limitation can be mitigated by using more clusters. In the extreme case of using the same number of clusters as training data samples, i.e., $I=\commentTHe{K}$, each test sample is approximated by the best fitting \commentTHb{\ac{NN} training sample}. 
%It is, however, not clear how this neighbour should be selected. 
Motivated by the \commentTHa{property} of manifolds being locally Euclidean, the \commentTHa{squared} Euclidean distance
\begin{equation}
\commentTHc{{d_{\text{euc}}(\underline{\boldsymbol{w}}_\tau, \underline{\boldsymbol{w}}_\kappa^{\text{tr}}) = ||\underline{\boldsymbol{w}}_\tau - \underline{\boldsymbol{w}}_\kappa^{\text{tr}}||^2}}
\label{eq:euclidean_dist_metric}
\end{equation}
%
% to evaluate the \ac{NN} approximation approach. The respective de
in between the ground-truth \commentTHc{\ac{AIR} vector} $\underline{\boldsymbol{w}}_\tau$ and the training data samples \commentTHc{$\underline{\boldsymbol{w}}_\kappa^{\text{tr}}$} is used to select the \commentTHc{\acp{NN}.}
The respective \ac{NN} approximation attains an average system mismatch of \makebox{$\commentTHe{\bar{\Upsilon}}=-6.7$} dB (cf.~\commentTHc{\textit{\ac{KNN} Proj.}} at subspace dimension \commentTHd{$D_i=0$} in Fig.~\ref{fig:subspace_system_mismatch}). We suspect that \commentTHc{the \ac{NN} model generalizes poorly to \acp{AIR} in between the training samples as the respective subspace is zero-dimensional, i.e., condensed to a single element.} 
%the model generalizes poorly \commentTHa{to \acp{AIR} in between the training samples}. 
To remedy this limitation we suggest to remove the condition of non-overlapping local training data sets $\mathcal{U}_i$ as used in \cite{9231543}.
% As this resulted only in poor performance we suggest, in comparison to \cite{mlsp_paper}, to remove the condition of non-overlapping local training data sets $\mathcal{U}_i$. 
% Thus, allows to use $G_I$ \acp{NN} to the training sample to compute the local affine subspace representative $\mathcal{M}_i$. Furthermore, one can extend this approach to use the $G_i$ \acp{NN} from the current test sample, instead of the closest training sample, to compute the local training data set $\mathcal{U}_i$.
% compute the local training data set $\mathcal{U}_i$ from the \ac{KNN} to the current test sample instead of the closest training sample.
In particular we propose to estimate the affine subspace parameters from the \commentTHe{$K_\tau$} \ac{NN} training samples \commentTHc{$\underline{\boldsymbol{w}}_\kappa^{\text{tr}}$} to the current \commentTHc{grond-truth} \ac{AIR} vector \commentTHd{$\underline{\boldsymbol{w}}_\tau$}. \commentTHd{Note that we index the subspace-related parameters by $\tau$ in the {following} to stress the dependence on the ground-truth \ac{AIR} vector.} The corresponding projection suggests a reconstruction of $\underline{\boldsymbol{w}}_\tau$ from its surrounding \commentTHe{$K_\tau$} training samples. \commentTHc{Note that if the subspace dimension is chosen as \commentTHe{$D_\tau=K_\tau-1$, with \commentTHe{$K_\tau$} being the number of neighbours}, the projection can be computed directly from the training samples (\textit{\ac{KNN} Proj.}) as shown in the sequel.}
%can be computed without any eigenvalue decomposition of the local covariance matrix \commentTHa{as shown in the sequel}. 
\commentTHd{Note that one degree of freedom is required for computing the local offset vector $ \bar{\underline{\boldsymbol{w}}}_\tau$ (cf.~Eq.~\eqref{eq:defProj}). 
%and we assume linearly independent training \ac{AIR} vectors. 
The projection matrix ${\underline{\boldsymbol{L}}_\tau}$ is obtained without eigenvalue decomposition of the local covariance matrix by choosing}
\begin{equation}
\commentTHd{
	\underline{\boldsymbol{V}}_\tau = \begin{bmatrix}
		\underline{\boldsymbol{w}}_1^{\text{tr}} - \bar{\underline{\boldsymbol{w}}}_{\tau} & \dots & \underline{\boldsymbol{w}}_{K_\tau-1}^{\text{tr}} - \bar{\underline{\boldsymbol{w}}}_{\tau}\end{bmatrix}}
	\label{eq:knn_basis_comp}
\end{equation}
%
% TODO: check this in code again
%
in Eq.~\eqref{eq:defProj}. \commentTHc{The \commentTHe{corresponding} \textit{\ac{KNN} Proj.} algorithm is} evaluated in Fig.~\ref{fig:subspace_system_mismatch}. The \commentTHe{respective $K_\tau$ training \acp{AIR}} are computed based on the \commentTHd{squared} Euclidean distance \eqref{eq:euclidean_dist_metric} w.r.t. the true \ac{AIR}. We observe that\commentTHd{, for an equivalent subspace dimension $D_i=D_\tau$,} the proposed method achieves a much lower average system mismatch \commentTHe{$\bar{\Upsilon}$} in comparison to both the global and the mixture approach. Furthermore, due to removing the condition of \commentTHd{disjoint} local training data sets, the \commentTHc{\ac{KNN}-based projection} achieves the benchmark performance \commentTHc{(cf.~\commentTHc{\textit{Oracle GT}})} at a much lower subspace dimension in comparison to the mixture approach \commentTHc{(cf.~\commentTHc{\textit{Mixture Proj.}})}. \commentTHd{This property is beneficial as the orthogonal space to the affine subspace is treated as noisy part of an \ac{AF} estimate in the following} \commentTHe{(cf.~Sec.~\ref{sec:softSubProj}).}
\section{Acoustic Impulse Response Denoising}
% \label{sec:airDenAdSubProj}
\label{sec:kf}
%\begin{itemize}
%	\item introduction
%\end{itemize}
% In this section we will introduce the proposed \ac{OSASI} algorithm which suggests a frequency-dependent convex combination of a \ac{KF} estimate and its projection on an adaptive learned \ac{AIR} subspace model. 
In this section, we introduce the proposed \ac{OSASI} algorithm which fuses a state-of-the-art \ac{KF} adaptation with an \commentTHc{adaptive} \ac{AIR} subspace model. The subspace parameters are estimated from the \commentTHe{\ac{NN}} training samples (cf.~Sec.~\ref{sec:air_mod_par_est}). For computing the closest neighbours we propose a novel distance which takes the state uncertainty of the \ac{KF} into account. Finally, we describe a probabilistically motivated frequency-dependent convex combination of the \ac{KF} estimate and its projection onto the affine subspace which \commentTHe{improves the performance of the baseline \ac{KF}.} 

%The probabilistically motivated combination of the \ac{AF} estimate and its projection on the affine subspace allows for faster convergence in noisy applications.
%
%
% DSSP (Denoising by Soft Subspace Projection)
% \textcolor{red}{Think of a better word than soft!}

\subsection{Kalman Filter-based Acoustic Impulse Response Estimation}
\label{sec:kf_air_est}
%
%\begin{itemize}
%	\item introduce the classical Kalman filter
%	\item shortly discuss the interpretation of the step size
%\end{itemize}
%
The \ac{DFT}-domain \ac{KF} \cite{enzner_frequency-domain_2006, malik_recursive_2011} approach to \ac{OSASI} suggests a probabilistic inference of the latent \commentTHe{\ac{ATF}} vector $\boldsymbol{w}_\tau$. For this\commentTHc{,} the conditional \ac{PDF} of $\boldsymbol{w}_\tau$, given the \commentTHa{current and the} preceding observations \makebox{${\boldsymbol{Y}}_{1:\tau} = \begin{bmatrix}{\boldsymbol{y}}_{1}, & \dots, & {\boldsymbol{y}}_{\tau}\end{bmatrix}$}, is modeled by \cite{malik_recursive_2011}
\begin{equation}
p({\boldsymbol{w}}_{\tau} | {\boldsymbol{Y}}_{1:\tau} ) = \mathcal{N}_c \left({\boldsymbol{w}}_{\tau}|\kfMean_{\tau}, {\bP}_{\tau}   \right)
\label{eq:data_likelihood}
\end{equation}
with the \commentTHe{\ac{ATF}} \commentTHa{mean} vector 
\begin{equation}
	\kfMean_{\tau} = \begin{bmatrix} \left(\kfMean_{1,\tau}\right)^{\trans} & \dots & \left(\kfMean_{B,\tau}\right)^{\trans}
	\end{bmatrix}^{\trans} \in \mathbb{C}^{MB}
\end{equation}
and \commentTHe{the} state uncertainty matrix 
\begin{equation}
	{\bP}_{\tau} = \begin{bmatrix} {\bP}_{11, \tau} & \dots & {\bP}_{1B, \tau}\\
	\vdots & \ddots & \vdots \\
	{\bP}_{B1, \tau} & \dots & {\bP}_{BB, \tau} 
	\end{bmatrix} \in \mathbb{C}^{BM \times BM}.
	\label{eq:state_unc_cov_mat}
\end{equation}
Due to the linear Gaussian \ac{DFT}-domain state transition model \eqref{eq:stateTransMod} and observation model \eqref{eq:fd_obs_eq}, a closed form recursive update of the likelihood \eqref{eq:data_likelihood} is given by the \ac{KF} equations \cite{bishop2007}. 
% Furthermore, by assuming diagonal state uncertainty matrices ${\bP}_{ij, \tau}$, a
In particular by assuming a \commentTHc{diagonal structure for the submatrices ${\bP}_{ij, \tau}$ \makebox{($i,j\in \{ 1,\dots,B\}$)} of the state uncertainty matrix ${\bP}_{\tau}$ computationally efficient update rules are obtained \cite{malik_recursive_2011}:}
%
%\begin{align} % TODO: check this again and maybe take A out in prediction
%\label{eq:eStep}
%%\hat{\boldsymbol{w}}^{+}_{b,\tau - 1} &= \hat{\boldsymbol{w}}_{b,\tau - 1} \notag \\ % maybe take A out here
%% &	\widehat{\boldsymbol{d}}_{\tau}=\sum_{b=0}^{B-1} {\C}_{b,\tau} \hat{\boldsymbol{w}}_{b,\tau - 1} \notag \\
%% &	{\e}_{\tau}^{+} 					=  {\y}_{\tau} - \widehat{\boldsymbol{d}}_{\tau} \notag\\
%&	{\e}_{\tau}^{+} 					=  {\y}_{\tau} -  {\C}_{\tau} \kfMean_{\tau - 1} \notag\\
%&{\bP}^{+}_{b,\tau - 1} 				= \commentTHa{a}^2 ~ {\bP}_{b,\tau - 1} + {\boldsymbol{\Psi}}^{\Delta\text{W}}_{b,\tau} \\
%% &\boldsymbol{\Lambda}_{b,\tau} 		= {\bP}^{+}_{b,\tau - 1} \left(\sum_{b=0}^{B-1} {\X}_{\tau-b} {\bP}^{+}_{b,\tau-1} {\X}_{\tau-b}^{\herm}  + \frac{M}{R} {\boldsymbol{\Psi}}^{\text{I}}_{\tau}\right)^{-1} \notag\\
%&\boldsymbol{\Lambda}_{b,\tau} 		= {\bP}^{+}_{b,\tau - 1} \left({\sum_{\tilde{b}=1}^{B} {\X}_{\tilde{b},\tau} {\bP}^{+}_{\tilde{b},\tau-1} {\X}_{\tilde{b},\tau}^{\herm}} + \frac{M}{R} {\boldsymbol{\Psi}}^{\text{N}}_{\tau}\right)^{-1} \notag\\
%% \hat{\boldsymbol{w}}_{b,\tau}	 	& = \hat{\boldsymbol{w}}^{+}_{b,\tau - 1} + \boldsymbol{G} \boldsymbol{\Lambda}_{b,\tau} {\X}_{b,\tau}^{\herm} {\e}_{\tau}^{+} \notag\\
%&\kfMean_{b,\tau}	 	 = \kfMean_{b,\tau - 1} + \boldsymbol{G} \boldsymbol{\Lambda}_{b,\tau} {\X}_{b,\tau}^{\herm} {\e}_{\tau}^{+} \notag\\
%&{\bP}_{b,\tau} 						 =  \left({\I}_M - \frac{R}{M} \boldsymbol{\Lambda}_{b,\tau} {\X}_{b,\tau}^{\herm} {\X}_{b,\tau}\right) {\bP}^{+}_{b,\tau-1}. \notag
%\end{align}
%
\begin{align} % TODO: check this again and maybe take A out in prediction
% \label{eq:adsdf} % \label{eq:eStep}
%\hat{\boldsymbol{w}}^{+}_{b,\tau - 1} &= \hat{\boldsymbol{w}}_{b,\tau - 1} \notag \\ % maybe take A out here
% &	\widehat{\boldsymbol{d}}_{\tau}=\sum_{b=0}^{B-1} {\C}_{b,\tau} \hat{\boldsymbol{w}}_{b,\tau - 1} \notag \\
% &	{\e}_{\tau}^{+} 					=  {\y}_{\tau} - \widehat{\boldsymbol{d}}_{\tau} \notag\\
%&\hat{\boldsymbol{d}}_\tau 			=   {\C}_{\tau} \kfMean_{\tau - 1} \approx  \commentTHd{A} {\C}_{\tau} \kfMean_{\tau - 1} \label{eq:echo_est} \\
% &	{\e}_{\tau}^{+} 					=  {\y}_{\tau} -  \hat{\boldsymbol{d}}_\tau
&	{\e}_{\tau}^{+}  = {\y}_{\tau} -  {\C}_{\tau} \kfMean_{\tau - 1} \approx {\y}_{\tau} - \commentTHd{A}~ {\C}_{\tau} \kfMean_{\tau - 1} \label{eq:error_comp} \\
&{\bP}^{+}_{i j,\tau - 1} 				= \commentTHd{A}^2 ~ {\bP}_{i j,\tau - 1} + {\boldsymbol{\Psi}}^{\Delta\text{W}}_{ij,\tau} \label{eq:state_cov_pred}  \\
% &\boldsymbol{\Lambda}_{b,\tau} 		= {\bP}^{+}_{b,\tau - 1} \left(\sum_{b=0}^{B-1} {\X}_{\tau-b} {\bP}^{+}_{b,\tau-1} {\X}_{\tau-b}^{\herm}  + \frac{M}{R} {\boldsymbol{\Psi}}^{\text{I}}_{\tau}\right)^{-1} \notag\\
&\boldsymbol{D}_\tau = {\sum_{i,j=1}^{B} {\X}_{i,\tau} {\bP}^{+}_{ij,\tau-1} {\X}_{j,\tau}^{\herm}} + \frac{M}{R} {\boldsymbol{\Psi}}^{\text{N}}_{\tau}  \\
&\boldsymbol{\Lambda}_{i,\tau}		=  \sum_{j=1}^{B} \left( {\bP}^{+}_{ij,\tau - 1}  {\X}_{j,\tau}^{\herm}   \right)\boldsymbol{D}_\tau^{-1} \label{eq:obs_pow} \\
% \hat{\boldsymbol{w}}_{b,\tau}	 	& = \hat{\boldsymbol{w}}^{+}_{b,\tau - 1} + \boldsymbol{G} \boldsymbol{\Lambda}_{b,\tau} {\X}_{b,\tau}^{\herm} {\e}_{\tau}^{+} \notag\\
&\kfMean_{i,\tau}	 	 = \kfMean_{i,\tau - 1} + \boldsymbol{G} \boldsymbol{\Lambda}_{i,\tau} {\e}_{\tau}^{+} \label{eq:kf_update} \\
&{\bP}_{ij,\tau} 						 =  {\bP}^{+}_{ij,\tau-1} - \frac{R}{M} \boldsymbol{K}_{i,\tau} \sum_{l=1}^B {\X}_{l,\tau} {\bP}^{+}_{lj,\tau-1}. \label{eq:state_cov_upd} 
\end{align}
%
%\textcolor{violet}{Check KF Equations here and in EUSIPCO PAPER!!!!!!!! Also notation of state uncertainty with ${\bP}_{b,\tau}$ and ${\bP}_{bb,\tau}$ is not consistent!!!! Attention change also the notaiton for the projeciton matrix !!!! TODO: check if this is exactly what you implemented !!! do not use $b_1$ and $b_2$  \vspace*{1cm}}
%
Note that\commentTHc{, in contrast to \cite{malik_recursive_2011},} we introduced a gradient constraint matrix $\boldsymbol{G} = \boldsymbol{F}_M \boldsymbol{Q}_2 \boldsymbol{Q}_2^{\text{T}} \boldsymbol{F}_M^{-1}$ to ensure a \commentTHc{zero-padded time-domain} \ac{AIR} vector \cite{buchner_generalized_2005} and set the state transition factor $\commentTHd{A}$ to one in the \commentTHb{prior error computation \eqref{eq:error_comp} \cite{kuech_state-space_2014}}. The resulting update rules \commentTHb{\eqref{eq:error_comp}~-~\eqref{eq:state_cov_upd}} can be interpreted as an extension of the classical \ac{FDAF} \cite{haykin_2002} to multiple excitation signals including an adaptive frequency-dependent step size \commentTHb{embedded in the Kalman gain matrix $\boldsymbol{\Lambda}_{i,\tau}$}. % matrix $\boldsymbol{\Lambda}_{b,\tau} \in \mathbb{C}^{M \times M}$ with $b\in \{ 1,\dots,B\}$.

The required process and observation noise covariance matrices 
\begin{align}
%	{\boldsymbol{\Psi}}_{\tau}^{\text{W}} &= \lambda_W ~{{\boldsymbol{\Psi}}_{\tau-1}^{\text{W}}}  + (1-\lambda_W) ~\kfMean_{\tau} \left(\kfMean_{\tau}\right)^{\text{H}} \label{eq:mStepFiltPowEst} \\
{\boldsymbol{\Psi}}_{\tau}^{\Delta\text{W}} &= (1-\commentTHd{A}^2)~{\boldsymbol{\Psi}}_{\tau}^{\text{W}} \label{eq:mStepStateNoise} \\
 {\boldsymbol{\Psi}}_{\tau}^{\text{N}} & =\lambda_{\text{N}}~ {\boldsymbol{\Psi}}_{\tau-1}^{\text{N}}  + (1-\lambda_{\text{N}}) ~ {\boldsymbol{e}}_{\tau}^+ {\left({\boldsymbol{e}}_{\tau}^+ \right)}^{\text{H}}\label{eq:mStepObsNoise} 
\end{align}
are estimated from the observed microphone signals using the estimated \ac{ATF} power
\begin{equation}
		\label{eq:mStepFiltPowEst}
	{\boldsymbol{\Psi}}_{\tau}^{\text{W}} =\lambda_{\text{W}} ~{{\boldsymbol{\Psi}}_{\tau-1}^{\text{W}}}  + (1-\lambda_{\text{W}}) ~\kfMean_{\tau-1} \left(\kfMean_{\tau-1}\right)^{\text{H}} 
\end{equation}
and the recursive averaging factors \commentTHa{$\lambda_{\text{W}}$ and $\lambda_{\text{N}}$} \makebox{\cite{kuech_state-space_2014, yang_frequency-domain_2017,franzen_improved_2019}}.
%
%\clearpage
%\newpage

\subsection{Adaptive Subspace Tracking}
\label{sec:adSubProj}
In \commentTHb{Sec.~\ref{sec:air_mod_par_est}} we described the idea of learning an affine subspace \commentTHe{$\mathcal{M}_\tau$} \commentTHc{for} the current test sample \commentTHe{$\underline{\boldsymbol{w}}_\tau$} from the surrounding \commentTHe{$K_\tau$ \acp{NN}} in the training data. This approach is straightforwardly extended to the \ac{OSASI} application by computing the affine subspace $\mathcal{M}_\tau$ based on the \commentTHe{\acp{NN}} w.r.t to the current \ac{AF} estimate $\kfMean_{\tau}$ (cf.~Eq.~\eqref{eq:data_likelihood}). 
%\commentTHb{Note that we used here the index $\tau$ to stress the dependence on the \commentTHd{current} \ac{AF} estimate $\boldsymbol{w}^{\text{kf}}_\tau$.}
%This renders however the problem that due to the online estimation the current estimate $\hat{\boldsymbol{w}}_{\tau}$ does not need to be close to the optimum
% It is however not yet clear how the neighbours should be selected. In Sec.~\ref{sec:knn_aff_subspace} we suggested to use the Euclidean distance \eqref{eq:euclidean_dist_metric} which was motivated by the assumptions of manifolds being locally Euclidean.
We now \commentTHd{discuss} the question if there are better choices than the simple \commentTHd{squared} Euclidean distance \eqref{eq:euclidean_dist_metric} for computing the closest neighbours in the training data set. For this we exploit the probabilistic \ac{KF} model \eqref{eq:data_likelihood} which renders \commentTHc{an uncertainty measure ${\bP}_{\tau}$ of the current mean estimate $\kfMean_{\tau}$}. We suggest the \commentTHc{negated} likelihood of the training data samples given the \ac{KF} estimate \commentTHb{(cf.~Eq.~\ref{eq:data_likelihood})} as \commentTHa{squared} distance \commentTHa{measure}
\begin{align}
-\log ~	&p(\boldsymbol{w}_\kappa^{\text{tr}} | {\boldsymbol{Y}}_{1:\tau} ) \\
 &=- \log \mathcal{N}_c (\boldsymbol{w}_\kappa^{\text{tr}} | \kfMean_\tau, \boldsymbol{P}_\tau) \\
&  \stackrel{\text{c}}{=} \left(  \boldsymbol{w}_\kappa^{\text{tr}} - \kfMean_\tau \right)^{\text{H}} \boldsymbol{P}_\tau^{-1} \left(  \boldsymbol{w}_\kappa^{\text{tr}} - \kfMean_\tau \right) \label{eq:non_diag_weight_kf_dist} \\
&= d_{\text{kf}}(\boldsymbol{w}_\kappa^{\text{tr}}, \kfMean_\tau) .
\label{eq:prob_dist_measure}
\end{align}
Eq. \eqref{eq:prob_dist_measure} describes a frequency-dependent weighted \commentTHb{squared} Euclidean distance \commentTHe{for which} more reliable estimates, i.e., \commentTHc{those} having a lower state uncertainty, are more important, i.e., have a higher weight. Note that if the state uncertainty matrix $\boldsymbol{P}_\tau$ is chosen as identity matrix, as usually done in classical \ac{AF} algorithms, the probabilistic distance measure \eqref{eq:prob_dist_measure} simplifies to the \commentTHd{squared} Euclidean distance \eqref{eq:euclidean_dist_metric}. 
%
%\begin{itemize}
%	\item Explain KNN approach to estimate the subspace 
%	\item introduce the various distance measures
%	\begin{itemize}
%		\item Euclidean distance
%		\item state uncertainty of Kalman filter
%		\item attention only diagonal elements
%	\end{itemize}
%	\item Explain the extraction of the basis from the KNN and relate to local PCA
%\end{itemize}
%
%
%\begin{align}
% d(\boldsymbol{w}_\kappa^{\text{tr}}, \hat{\boldsymbol{w}}_\tau) &= \log	p(\boldsymbol{w}_\kappa^{\text{tr}}| \hat{\boldsymbol{w}}_\tau) \\
% & = \log \mathcal{N}_c (\boldsymbol{w}_\kappa^{\text{tr}} | \hat{\boldsymbol{w}}_\tau, \boldsymbol{P}_\tau) \\
% 	&  \stackrel{\text{c}}{=}  \left(  \boldsymbol{w}_\kappa^{\text{tr}} - \hat{\boldsymbol{w}}_\tau \right)^{\text{H}} \boldsymbol{P}_\tau^{-1} \left(  \boldsymbol{w}_\kappa^{\text{tr}} - \hat{\boldsymbol{w}}_\tau \right)
% 	\label{eq:prob_dist_measure}
%\end{align}
%
%Note that if the state uncertainty matrix $\boldsymbol{P}_\tau$ is chosen as identity matrix, the probabilistic distance measure \eqref{eq:prob_dist_measure} simplifies to the Euclidean distance $ d(\boldsymbol{w}_\kappa^{\text{tr}}, \hat{\boldsymbol{w}}_\tau) $ (cf.~Eq.~\eqref{eq:euclidean_dist_metric}).
%%
%\begin{align}
% d(\boldsymbol{w}_\kappa^{\text{tr}}, \hat{\boldsymbol{w}}_\tau) &= ||  \boldsymbol{w}_\kappa^{\text{tr}} - \hat{\boldsymbol{w}}_\tau ||_2^2
%\label{eq:euclidean_distance}
%\end{align}

\subsection{Soft Subspace Projection}
\label{sec:softSubProj}
%We now describe how the adaptively estimated affine subspace $\mathcal{M}_\tau$ (cfg.~Secs.~\ref{sec:aff_subspace_models}~and~\ref{sec:adSubProj}) is exploited to enhance the \ac{KF} estimate $\kfMean_\tau$.
%In Sec.~\ref{sec:airAnalysis} the assumption of an \ac{AIR} manifold, i.e., that all \acp{AIR} are confined to a structured subset of the estimation space $\mathbb{R}^Q$, has been introduced. Subsequently, we evaluated various affine subspace-based approaches to locally model the manifold. The modelling capability of the algorithms has been evaluated by the system mismatch of projecting a noise-free \ac{AIR} onto the models. 
%
In Sec.~\ref{sec:airAnalysis} the local affine subspace approximation of an \ac{AIR} manifold has been evaluated by the average system mismatch which results from a projection of a noise-free \ac{AIR} vector $\underline{\boldsymbol{w}}_\tau$ onto a local affine subspace. Here, the projective mapping \eqref{eq:projEstimate} sets \commentTHd{the coordinates} with little influence on the \ac{AIR} vector to zero, i.e., \commentTHd{yields a compressed description of the \ac{AIR} vectors}. This observation can now be exploited to denoise an \ac{AF} estimate ${\hat{\underline{\boldsymbol{w}}}}_\tau^{\text{kf}}$ by
\begin{equation}
  \hat{{\boldsymbol{{w}}}}_\tau^\text{kf,p} =  \left(\boldsymbol{I} \otimes \left( \boldsymbol{F}_M \boldsymbol{Q}_2 \right)\right) ~\commentTHe{f_{\mathcal{M}_\tau}} \left( {\hat{\underline{\boldsymbol{w}}}}_\tau^{\text{kf}} \right)
	% $f_{\mathcal{M}_\tau} (\cdot) $ (cf.~Eq.~
\end{equation}
with the time-domain \ac{AF} estimate 
\begin{equation}
	{\hat{\underline{\boldsymbol{w}}}}_\tau^{\text{kf}}  = \left( \boldsymbol{I} \otimes \left( \boldsymbol{Q}_2^{\trans} \boldsymbol{F}_M^{-1} \right)\right) ~ {\kfMean}_\tau \in  \mathbb{R}^Q.
\end{equation}
This projection \commentTHc{removes all components of the \ac{KF}-based} \ac{AIR} estimate \commentTHe{$\underline{\boldsymbol{w}}^{\text{kf}}_\tau$} which are not supported by the surrounding training data samples \commentTHd{spanning} the affine subspace $\mathcal{M}_\tau$. This is visualized in Fig.~\ref{fig:sof_proj_manifold_example}.
\begin{figure}[tb]
	%\documentclass[tikz,border=3.14mm]{standalone}
%\usepackage{pgfplots}
%\pgfplotsset{compat=1.15}
%\usetikzlibrary{calc}
%\begin{document}

\begin{tikzpicture}[]
\pgfplotsset{
	colormap={blackwhite}{gray(0cm)=(.4); gray(1cm)=(.9)}
}

\begin{axis}[
hide axis,
%view/h=45,
%xlabel = {$x$},
%ylabel = {$y$},
%zlabel = {$z$},
declare function={f(\x,\y)=10-(\x^2+\y^2);},		%f(x,y)
declare function={c_x(\t)=cos(\t)+1;},				% cx(t)
declare function={c_y(\t)=sin(\t)-1;},				% cy(t)
declare function={c_z(\t)=f(c_x(\t),c_y(\t));},zmax=9.25, samples=8]	% cz(t) = f(cx(t), cy(t))
\addplot3[surf,  domain=-1.7:1.7,domain y=-2:2,]{f(x,y)};
%\addplot3[black,opacity=1.0,variable=t,domain=0:360] ({c_x(t)},{c_y(t)},{c_z(t)});

\coordinate (center) at (150.0, 40, -0.0);

\draw [thick, ->] ($(center)+(0,0,0)$) -- ($(center)+(60,0,0)$) node [at end, right] {$[\underline{\boldsymbol{w}}_{(\cdot)}]_2$};
\draw [thick,->] ($(center)+(0,0,0)$) -- ($(center)+(0,-98,0)$); % node [at end, left,above] {$[\underline{\boldsymbol{w}}]_1$};
\draw [thick,->] ($(center)+(0,0,0)$) -- ($(center)+(0,0,160)$) node [at end, left] {$[\underline{\boldsymbol{w}}_{(\cdot)}]_3$};

\node (w1) at ($(center)+(-35,-70,20)$) {$[\underline{\boldsymbol{w}}_{(\cdot)}]_1$};

	%\draw[fill=white,fill opacity=0.8, tangent plane=at {(-.75,-.75,10)} with vectors {(1.5,0,0)} and {(0,1.5,0)}];
	% \draw[fill=white,fill opacity=0.8, tangent plane=at {(-1.25,-1.0,5)} with vectors {(1.5,0, 0)} and {(0,1,2.5)}];

%\draw[fill=white,fill opacity=0.8,tangent plane=at {(.5,-1.0,5)} with vectors {(1,2,-1.5)} and {(-1,1,2)}];

% \node at (-1.0,-.8, 5.4) {$\mathcal{M}_1$};

%	\node at (.0,0,10) {$\mathcal{M}_2$};

%	\draw [->, thick, red] (-0.4,-.8, 6.5) -- (0.3,-.8, 6.5);
%	\draw [->, thick, red] (-0.4,-.8, 6.5) -- (-.4,-.4, 7.5);

\end{axis}

	\coordinate (est) at (3.1, 5.5, 0.2);
\draw [circle, fill, red] ($(est)+(0,0,0)$) circle (2pt);
\node at ($(est)+(.45,.0,0)$) {\textcolor{black}{$\hat{\underline{\boldsymbol{w}}}_{\tau}^{\text{kf}}$}};

	\coordinate (proj) at (3.1, 4.0, 0.2);
	
\draw [dashed, thick, <->] ($(est)+(0,-.1, 0)$) -- (proj) node [midway, right] {$\hat{\underline{\boldsymbol{w}}}_{\tau}^{\text{den}}$};
\draw [circle, fill,orange] ($(est)+(0,-0.8, 0)$) circle (2pt);

	%\draw [circle, fill,blue] (2.0, 3.2, 0.0) circle (2pt);
	\draw [circle, fill,blue] (3.1, 3.9, 0.2) circle (2pt);
	%\draw [circle, fill,blue] (3.9, 4.2, 0.2) circle (2pt);

	% \node at (2.6, 3.1, 0.0) {\textcolor{black}{$\underline{\boldsymbol{w}}_{\tau-1}^{\text{p}}$}};
	\node at (3.6, 3.8, 0.2) {\textcolor{black}{$\hat{\underline{\boldsymbol{w}}}_{\tau}^{\text{kf,p}}$}};
	% \node at (4.5, 4.1, 0.2) {\textcolor{black}{$\underline{\boldsymbol{w}}_{\tau+1}^{\text{p}}$}};
	
		\node at (1.5,4.8, 0) {$\mathcal{M}_\tau$};
		\draw [dashed, thick] ($(1.3,4.6, 0)+(.2,0,0)$) to [out=-60,in=-180] ($(proj)+(-.1,-.3,0)$);

\draw [circle, fill,green] (2.6, 3.5, 0.2) circle (2pt);
\draw [circle, fill,green] (2.95, 3.5, 0.2) circle (2pt);
\draw [circle, fill,green] (2.75, 4.0, 0.2) circle (2pt);

\end{tikzpicture}

%\end{document}
	\caption{\commentTHc{The denoised estimate $\hat{\underline{\boldsymbol{w}}}_\tau^{\text{den}} \in \mathbb{R}^3$ is a convex combination of the \ac{KF}-based \ac{AIR} estimate $\hat{\underline{\boldsymbol{w}}}_\tau^{\text{kf}}$ and its projection $\hat{\underline{\boldsymbol{w}}}_\tau^{\text{kf,p}}$ onto the affine subspace $\mathcal{M}_\tau$. The \commentTHd{\acp{NN}} which span the subspace are visualized as green dots.}}
	\label{fig:sof_proj_manifold_example}
\end{figure}

However, the experiments in Sec.~\ref{sec:airAnalysis} showed also the \commentTHa{imperfection} of the local affine subspace models as \commentTHb{ground-truth} \acp{AIR} could not be perfectly reconstructed by the projection. This motivates the idea of modelling an uncertainty measure around the manifold, i.e., diluting the deterministic model. We suggest to model this uncertainty by a proper complex Gaussian \ac{PDF}
\begin{equation}
p({\boldsymbol{w}}_{\tau}|\commentTHa{\mathcal{M}_\tau}) = \mathcal{N}_c(\boldsymbol{w}_\tau| \hat{{\boldsymbol{{w}}}}_\tau^\text{kf,p} , \boldsymbol{\Psi}_{\mathcal{M}_\tau} )
\label{eq:prob_man_model}
\end{equation}
with the mean \commentTHb{vector} $ \hat{{\boldsymbol{{w}}}}_\tau^\text{kf,p} $ being the projection of $\kfMean_{\tau}$ onto the manifold and the covariance matrix \commentTHd{\makebox{$\boldsymbol{\Psi}_{\mathcal{M}_\tau} \in \mathbb{C}^{MB\times MB}$} expressing the uncertainty of the projection}. Subsequently, the probability of the \commentTHe{\ac{ATF}} vector $\boldsymbol{w}_\tau$ given the \ac{KF} likelihood \eqref{eq:data_likelihood} and the probabilistic manifold model \eqref{eq:prob_man_model} is given by 
\begin{align}
p(\boldsymbol{w}_{\tau}&|\boldsymbol{Y}_{1:\tau}, \mathcal{M}_{\tau}) =\\ &\mathcal{N}_c(\boldsymbol{w}_\tau | \kfMean_\tau, \boldsymbol{P}_\tau) ~ \mathcal{N}_c(\boldsymbol{w}_\tau| \hat{{\boldsymbol{{w}}}}_\tau^\text{kf,p}, \boldsymbol{\Psi}_{\mathcal{M}_\tau} )
%p(\boldsymbol{w}_{\tau}&|\boldsymbol{Y}_{1:\tau}, \mathcal{M}_{\tau}) \\
%&=  p(\boldsymbol{w}_{\tau}|\boldsymbol{Y}_{1:\tau}) ~  \commentTHa{p(\boldsymbol{w}_{\tau}|\mathcal{M}_{\tau})}\\
%&= \mathcal{N}_c(\boldsymbol{w}_\tau | \kfMean_\tau, \boldsymbol{P}_\tau) ~ \mathcal{N}_c(\boldsymbol{w}_\tau| \hat{{\boldsymbol{{w}}}}_\tau^\text{kf,p}, \boldsymbol{\Psi}_{\mathcal{M}_\tau} )
\label{eq:likelihood_model}
\end{align}
assuming $p(\boldsymbol{w}_{\tau}|\boldsymbol{Y}_{1:\tau}, \mathcal{M}_{\tau}) = p(\boldsymbol{w}_{\tau}|\boldsymbol{Y}_{1:\tau}) ~  \commentTHa{p(\boldsymbol{w}_{\tau}|\mathcal{M}_{\tau})}$. The \ac{ML} estimate \commentTHd{of ${\boldsymbol{w}}_\tau$ based on \eqref{eq:likelihood_model} is given by \cite{matrix_cookbook}}:
%
% TODO: check if this holds true for complex value pdfs too?
\begin{align}
\hat{\boldsymbol{w}}_\tau^{\text{den}} = \left(\boldsymbol{P}_\tau^{-1} +  \boldsymbol{\Psi}_{\mathcal{M}_\tau}^{-1} \right)^{-1} \left( \boldsymbol{P}_\tau^{-1} \kfMean_\tau + \boldsymbol{\Psi}_{\mathcal{M}_\tau}^{-1}  \hat{{\boldsymbol{{w}}}}_\tau^\text{kf,p}  \right).
\label{eq:soft_proj}
%\hat{\boldsymbol{w}}_\tau^{\text{den}} = \left(\boldsymbol{P}_\tau^{-1} +  \boldsymbol{\Psi}_{\mathcal{M}_\tau}^{-1} \right)^{-1} \left( \boldsymbol{P}_\tau^{-1} \hat{\boldsymbol{w}}_\tau + \boldsymbol{\Psi}_{\mathcal{M}_\tau}^{-1} {\boldsymbol{w}}_{\mathcal{M}_\tau}  \right)
\end{align}
If the \ac{KF} state uncertainty matrix $\boldsymbol{P}_\tau$ and the prior \commentTHa{covariance matrix} $\boldsymbol{\Psi}_{\mathcal{M}_\tau}$ are assumed to be diagonal, Eq.~\eqref{eq:soft_proj} simplifies to

\begin{equation}
\left[ \hat{\boldsymbol{w}}_{b,\tau}^{\text{den}}\right]_f = (1-\alpha_{b,\tau,f}) \left[ \kfMean_{b,\tau}\right]_f + \alpha_{b,\tau,f} \left[ \hat{\boldsymbol{w}}_{b,\tau}^{\text{kf,p}}\right]_f
\label{eq:convex_combination}
\end{equation}
with the \ac{ATF} index $b$, the frequency \commentTHb{bin} index $f$ and the convex combination weight
\begin{equation}
	\alpha_{b,\tau,f} = \frac{ \left[\boldsymbol{P}_{bb,\tau} \right]_{ff}  }{\left[\boldsymbol{P}_{bb,\tau} \right]_{ff} + \commentTHb{\left[\boldsymbol{\Psi}_{{\mathcal{M}_{\tau}},bb} \right]_{ff}}}.
	\label{eq:conv_comb_weights}
\end{equation}
Here, the $b$th block diagonal element of \makebox{$\boldsymbol{\Psi}_{\mathcal{M}_{\tau}}$} is denoted by \commentTHb{\makebox{$\boldsymbol{\Psi}_{\mathcal{M}_{\tau},bb} \in \mathbb{C}^{M \times M}$}}. Eq.~\eqref{eq:convex_combination} represents a frequency-dependent convex combination of the \ac{KF} estimate $\hat{\boldsymbol{w}}_{\tau}^{\text{kf}}$ \commentTHa{and} its projection $\hat{\boldsymbol{w}}_{\tau}^{\text{kf,p}}$ onto the manifold which is visualized in Fig.~\ref{fig:sof_proj_manifold_example}. The proposed approach favours the \ac{KF} estimate $\kfMean_{\tau}$ whenever the \ac{KF} state uncertainty $\boldsymbol{P}_{\tau}$ is smaller than the prior model uncertainty $\boldsymbol{\Psi}_{\mathcal{M}_{\tau}} $ \commentTHa{and vice versa}. 
%In contrast if the model uncertainty is smaller, the projected estimate $\hat{\boldsymbol{w}}_{\tau}^{\text{p}}$ is favoured.

We will conclude this section by \commentTHa{proposing a model for} the prior uncertainty $\boldsymbol{\Psi}_{\mathcal{M}_{\tau}}$. It is assumed that the model uncertainty 
\begin{equation}
	\boldsymbol{\Psi}_{\mathcal{M}_{\tau}} =  \frac{\beta_{\text{pr}}}{1-\commentTHd{A}^2}\boldsymbol{\Psi}^{\Delta \text{W}}_\tau
	\label{eq:model_prior}
\end{equation}
is a scaled version of the process noise covariance matrix $\boldsymbol{\Psi}^{\Delta \text{W}}_\tau $ in the Markov model \eqref{eq:stateTransMod} with $\beta_{\text{pr}} > 0$ being a hyperparameter. \commentTHc{This is motivated by the intrinsic \commentTHe{\ac{ATF}} variability which is assumed to be proportional \commentTHe{to} the process noise power.} % This model assumes that the uncertainty around the manifold is a scaled version of the that the intrinsic \ac{AIR} movement 
\subsection{Algorithmic Description}
\label{sec:alg_descr}
Alg.~\ref{alg:prop_alg_descr} provides a detailed algorithmic description of the proposed \ac{KF-ASP} algorithm for \ac{OSASI}. For each block of microphone observations $\underline{\boldsymbol{y}}_\tau$ and loudspeaker excitations $\underline{\boldsymbol{x}}_\tau$, the prior error $\boldsymbol{e}_\tau^+$ is computed by using the \commentTHe{\ac{ATF}} estimate \commentTHa{of the previous time step} $\kfMean_{\tau-1}$ \commentTHb{(cf.~Eq.~\eqref{eq:error_comp})}. Subsequently, the process noise and observation noise covariance matrices ${\boldsymbol{\Psi}}_{\tau}^{\Delta\text{W}}$ and ${\boldsymbol{\Psi}}_{\tau}^{\text{N}}$ are updated by \commentTHc{Eqs.~\eqref{eq:mStepStateNoise}~-~\eqref{eq:mStepFiltPowEst}}. Note that \commentTHc{for computational efficiency} only the diagonal elements are updated to ensure the submatrix diagonality of the state uncertainty matrix $\bP_{\tau}$ (cf.~Eq.~\eqref{eq:state_cov_upd}). Afterwards, \commentTHd{the} posterior mean \commentTHa{vector} $\kfMean_\tau$ and \commentTHd{the} state uncertainty matrix $\bP_{\tau}$ are updated by the \ac{KF} Eqs. \commentTHa{\eqref{eq:state_cov_pred}-\eqref{eq:state_cov_upd}}. The \commentTHe{$K_\tau$ \acp{NN}} to the posterior mean \commentTHa{vector} $\kfMean_\tau$ are computed by the respective distance metric, \commentTHd{i.e.}, $d_{\text{euc}}(\cdot, \cdot)$ or $d_{\text{kf}}(\cdot, \cdot)$ (cf.~Eqs. \eqref{eq:euclidean_dist_metric}, \eqref{eq:prob_dist_measure}). \commentTHc{For} an efficient computation of the inverse in \commentTHd{Eq.~\eqref{eq:non_diag_weight_kf_dist}}, we only use the \commentTHc{elements of the main diagonal} of the state uncertainty matrix $\bP_{\tau}$. This \commentTHa{reduces} the computation of a \commentTHa{full} matrix inverse to a \commentTHa{element-wise} scalar inversion. The corresponding \commentTHe{$K_\tau$} closest samples \commentTHa{w.r.t. $d_{\text{euc}}$ or $d_{\text{kf}}$} are used as local training data \commentTHd{set} \commentTHd{$\mathcal{U}_\tau$} to compute the associated subspace offset $\bar{\underline{\boldsymbol{w}}}_\tau$ and projection matrix \commentTHd{$\underline{\boldsymbol{L}}_\tau$} by Eqs. \eqref{eq:mean_comp} and \eqref{eq:defProj}, respectively. \commentTHb{Because of the chosen} subspace dimension \commentTHe{$D_\tau=K_\tau-1$}, the basis matrix $\underline{\boldsymbol{V}}_\tau$ Eq.~\eqref{eq:knn_basis_comp} can be used in Eq.~\eqref{eq:defProj}.
% Note that due to choice of the subspace dimension $D_i$ being the number of neighbours $K_i$ minus one, i.e., $D_i=K_i-1$, the projection matrix can be computed directly by Eq.~\eqref{eq:defProj} with \makebox{$\underline{\boldsymbol{V}}_i = \begin{bmatrix} \underline{\boldsymbol{w}}_1^{\text{tr}} -\bar{\underline{\boldsymbol{w}}}_{\tau}& \dots & \underline{\boldsymbol{w}}_{K_i-1}^{\text{tr}}-\bar{\underline{\boldsymbol{w}}}_{\tau}\end{bmatrix}$}. 
The \ac{KF} estimate $\kfMean_\tau$ is then projected onto the affine subspace $\mathcal{M}_\tau$ by Eq.~\eqref{eq:projEstimate}. Subsequently, the convex combination weights $\alpha_{b,\tau,f}$ (cf.~Eq.~\eqref{eq:conv_comb_weights}) are used to fuse the \ac{KF} estimate $\kfMean_\tau$ and its projection $\hat{\boldsymbol{w}}_\tau^{\text{kf,p}}$. Finally, the respective denoised estimated \commentTHa{filter vector} $\hat{\boldsymbol{w}}_\tau^{\text{den}}$ is used as posterior mean \commentTHa{vector} $\kfMean_\tau$ of the \ac{KF}.

% \textcolor{red}{commit and then write posterior mean with kf subscript!}
%
\begin{algorithm}[tb] %[tb]
	\caption{\commentTHa{Proposed \ac{KF-ASP}-based \ac{AIR} estimation for one signal block $\underline{\boldsymbol{y}}_\tau$.}} % (\textcolor{red}{Check that this is exactly what is implemented in algorithm})
	\label{alg:prop_alg_descr}
	\begin{algorithmic}
	%	\For{$\tau=1,\dots,T$}
			\State Compute prior error $\boldsymbol{e}_\tau^+$ by Eq.~\eqref{eq:error_comp}
			\State Estimate ${\boldsymbol{\Psi}}_{\tau}^{\Delta\text{W}}$ and ${\boldsymbol{\Psi}}_{\tau}^{\text{N}}$ by \commentTHc{Eqs.~\eqref{eq:mStepStateNoise}-\eqref{eq:mStepFiltPowEst}}.
			\State Update $\kfMean_\tau$ and $\bP_{\tau}$ by \commentTHa{Eqs. \eqref{eq:state_cov_pred}-\eqref{eq:state_cov_upd}}
			\State Compute \ac{NN} training samples by distance metric $d_{(\cdot)}(\cdot, \cdot)$
			\State Compute subspace offset $\bar{\underline{\boldsymbol{w}}}_\tau$ by Eq. \eqref{eq:mean_comp} 
			\State Compute projection matrix $\underline{\boldsymbol{L}}_\tau$ by Eq.~\eqref{eq:defProj}
			\State Project \ac{KF} estimate onto affine subspace $\mathcal{M}_\tau$ by Eq.~\eqref{eq:projEstimate}
			\State Compute convex combination weights $\alpha_{b,\tau,f}$ by Eq.~\eqref{eq:conv_comb_weights}
			\State Compute denoised \ac{KF} estimate $\hat{\boldsymbol{w}}_\tau^{\text{den}}$ by Eq. \eqref{eq:convex_combination}
			\State Assign denoised estimate to \ac{KF}: $\kfMean_\tau \gets \hat{\boldsymbol{w}}_\tau^{\text{den}}$ 
	%	\EndFor
	\end{algorithmic}
\end{algorithm}

%\newpage
%\clearpage

\section{Experiments}
\label{sec:experiments}
We now evaluate the proposed algorithm for a typical \ac{MISO} acoustic system identification scenario. The acoustic scene is characterized by a loudspeaker array comprising $B=2$ elements with a spacing of $10$ cm and a single microphone. \commentTHc{Loudspeakers and microphone} are located in a room of dimensions \makebox{$[6~\text{m},~5~\text{m},~3.5~\text{m}]$} and a reverberation time of \makebox{$T_{60}=0.3$ s}. While the position of the loudspeaker array is kept fixed at \makebox{$[3.0~\text{m},~2.0~\text{m},~1.2~\text{m}]$}, the microphone position varies in a volumetric segment of a sphere with a radius \commentTHc{in the range} of \makebox{$r\in[1.2 ~\text{m},~1.4~\text{m}]$}, an \commentTHc{azimuthal angular} range \makebox{$\theta \in [45 \degree,~135 \degree]$} and elevation angle \commentTHc{in the range of} \makebox{$\phi \in [-5 \degree,~ 40 \degree]$} relative to the center of the loudspeaker array. All \acp{AIR} of length $W=4096$ have been simulated \commentTHc{using} the image method \cite{allen1979image, habets2010room} with a maximum reflection order and a sampling frequency of $f_s=8$ kHz. For each experiment the noise-free observation $\underline{\boldsymbol{d}}_{\tau}$ is simulated by convolving \commentTHe{the \acp{AIR} in} $\underline{\boldsymbol{w}}_\tau$, corresponding to a random observation position in the sphere segment, with a randomly chosen excitation signal $\underline{\boldsymbol{x}}_\tau$. Here, we consider two types of excitation signals: \commentTHe{spatially uncorrelated stationary \ac{WGN}} and speech. The speech signals are taken from a subset of the UWNU database \cite{uwnu_corpus} \commentTHd{which comprises} $15$ different speakers. \commentTHc{For the speech-excited scenarios we consider in addition to playing independent speech signals \commentTHe{at each loudspeaker} also a teleconferencing setup with strong correlation between the loudspeaker signals.}
% For the speech scenario we simulate two scen
%Subsequently, the noise-free observation $\underline{\boldsymbol{d}}_{\tau}$ is superimposed by noise signals $\underline{\boldsymbol{n}}_{\tau}$ which are scaled according to a desired \ac{SNR}.
\commentTHc{The additive noise signal $\underline{\boldsymbol{n}}_{\tau}$ is composed of \commentTHe{an interfering} \ac{WGN} component $\underline{\boldsymbol{n}}_{\text{wgn},\tau}$ and a nonstationary speech component $\underline{\boldsymbol{n}}_{\text{sp},\tau}$, i.e., $\underline{\boldsymbol{n}}_\tau=\underline{\boldsymbol{n}}_{\text{wgn},\tau}+\underline{\boldsymbol{n}}_{\text{sp},\tau}$. The variances of the noise signals $\underline{\boldsymbol{n}}_{\text{wgn},\tau}$ and $\underline{\boldsymbol{n}}_{\text{sp},\tau}$ are prescribed by $\text{SNR}_{\text{wgn}}$ and $\text{SNR}_{\text{sp}}$, respectively. The \commentTHd{set of} interfering speech signals \commentTHd{consists} of $15$ additional talkers from \cite{uwnu_corpus}.}

To evaluate the performance of the proposed algorithm we use the system mismatch $\Upsilon_\tau$ (cf.~Eq.~\eqref{eq:systMisDef1}) and the \ac{ERLE}
% maybe describe by expectation and discus that temporal measures is achieved by temporal recursive averaging and that scalar is obtained by arithmetic aveage
\begin{align}
{\mathcal{E}}_{\tau} = 10 \log_{10} \frac{\mathbb{E}\left[||\underline{\boldsymbol{d}}_{\tau}||^2\right] }{\mathbb{E}\left[||\underline{\boldsymbol{d}}_{\tau}-\widehat{\boldsymbol{\underline{d}}}_{\tau}||^2\right]}
\label{eq:erle_def}
\end{align}
\commentTHd{with the noise-free observation estimate $\widehat{\boldsymbol{d}}_\tau = {\C}_{\tau} \kfMean_{\tau - 1}$ (cf.~Eq.~\eqref{eq:error_comp})}.
Note that\commentTHc{, in contrast to the system mismatch $\Upsilon_\tau$,} the \ac{ERLE} ${\mathcal{E}}_{\tau}$ represents a signal-dependent performance measure \commentTHc{and, thus, is of} particular interest for signal cancellation applications. The expectation operator $\mathbb{E}[\cdot]$ in Eq.~\eqref{eq:erle_def} is \commentTHc{here} approximated by recursive averaging over time. To allow for more general conclusions, the system mismatch $\Upsilon_\tau$ and the \ac{ERLE} $\mathcal{E}_\tau$ are averaged over $50$ \commentTHc{trials of the random experiment with} randomly chosen excitation signals, random microphone positions and random interfering noise signals. The respective averaged performance measures are denoted by an overbar $\bar{(\cdot)}$.
\begin{figure}[t]
	\includegraphics[]{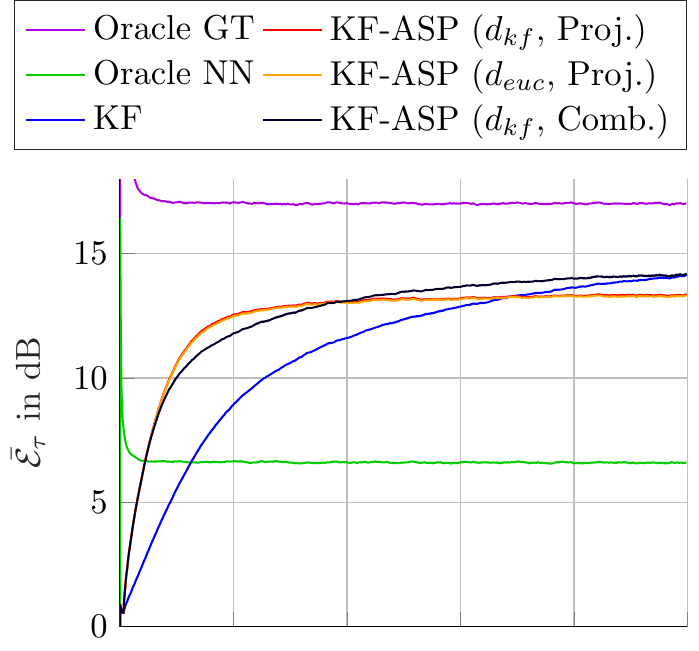}
	\hspace*{-.0cm}
	\includegraphics[]{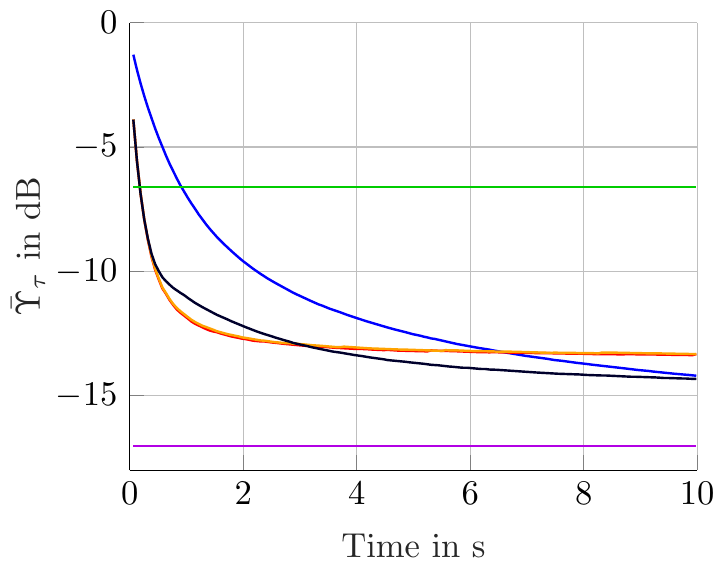}
	\caption{\commentTHc{Evaluation of the proposed KF-ASP-based algorithmic variants for a system identification application ($\text{SNR}_{\text{wgn}}=0$ dB, $\text{SNR}_{\text{sp}}=\infty$ dB) with a spatially uncorrelated \commentTHe{\ac{WGN}} excitation signal.}}
	% \caption{rendering wgn lr0 enr 0 dt Inf numTrainSamples 5000 (maybe delete prior ext)}
	\label{fig:wgn_exc}
\end{figure}
\begin{figure} [t]
	\includegraphics[]{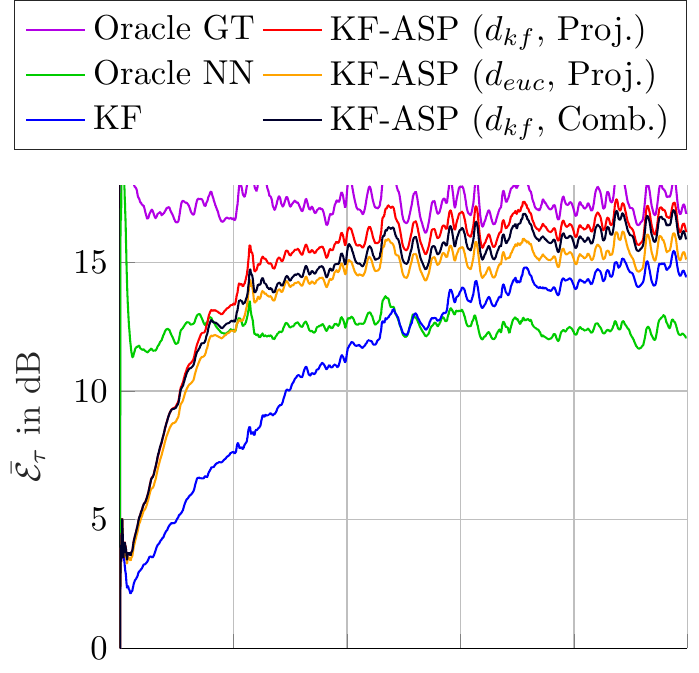}
	\hspace*{.2cm}
	\includegraphics[]{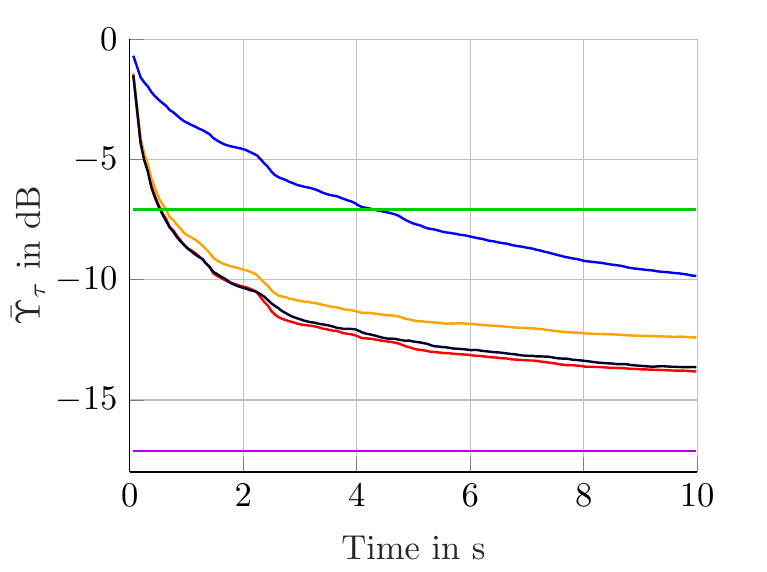}
	%\caption{rendering speech lr0 enr 5 dt 0 numTrainSamples 5000  (maybe delete prior ext)}
	% \caption{\commentTHc{Evaluation of the proposed affine subspace-based \ac{OSASI} algorithms for a scenario with two independent speech excitation signals. The noise-free observation is distorted by a mixture of a} \ac{WGN} signal at an \ac{SNR} of $5$ dB and a speech signal at an \ac{SNR} of $0$ dB.}
	\caption{\commentTHc{Evaluation of the proposed KF-ASP-based algorithmic variants for a system identification application ($\text{SNR}_{\text{wgn}}=5$ dB, $\text{SNR}_{\text{sp}}=0$ dB) with two independent speech excitation signals.}}
	\label{fig:speech_exc}
\end{figure}
\begin{figure} [h!]
	\hspace*{-0.25cm}
	\includegraphics[]{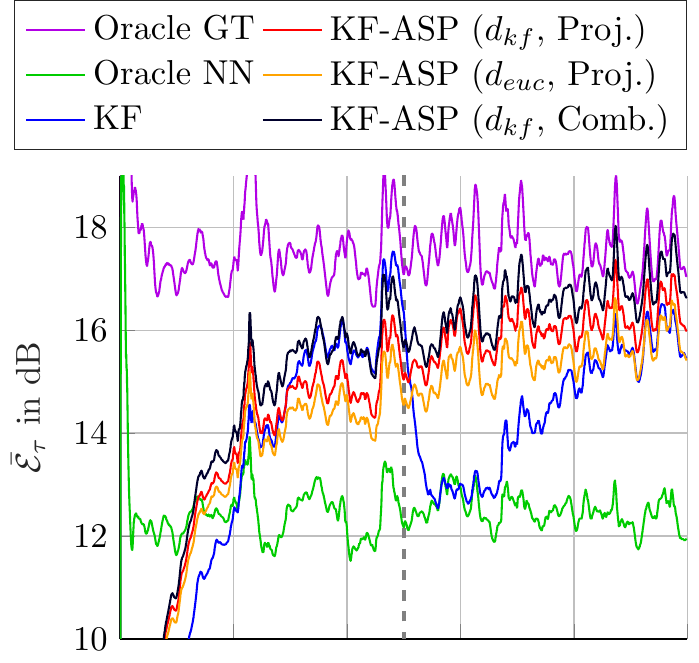}
	\includegraphics[]{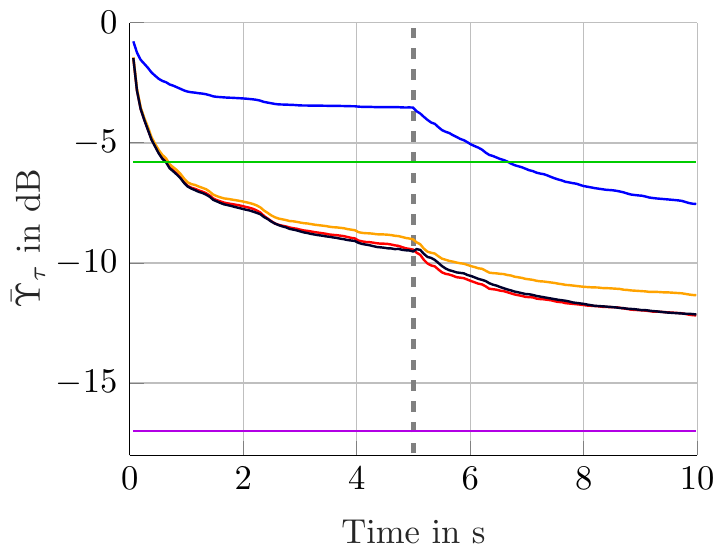}
	%	\caption{rendering speech lr1 enr 10 dt Inf numTrainSamples 5000 (\textcolor{red}{ylim from 10 to 20 dB})}
%	\caption{\commentTHc{Evaluation of the proposed affine subspace-based \ac{AEC} algorithm in a teleconferencing setup. Here, a change in the far-end speaker location is simulated after $5$ s. \ac{WGN} at an \ac{SNR} of $10$ dB is added to the noise-free microphone observations.}}
	\caption{\commentTHd{Evaluation of the proposed KF-ASP-based algorithmic variants for a teleconferencing setup ($\text{SNR}_{\text{wgn}}=10$ dB, $\text{SNR}_{\text{sp}}=\infty$ dB). After $5$ s the far-end speakers switch (indicated by a dashed gray line).}}
	% Here, a change in the far-end speaker location is simulated after $5$ s and .}
	\label{fig:non_unique_evaluation}
\end{figure}
\begin{figure} [t]
	% comparison of trainign sampel size for 1000
	%\setlength\fwidth{.7\columnwidth}
	% \input{images/erle_rendering_speech_lr0_enr_0_dt_Inf_numTrainSamples_1000.tikz}
	\includegraphics[]{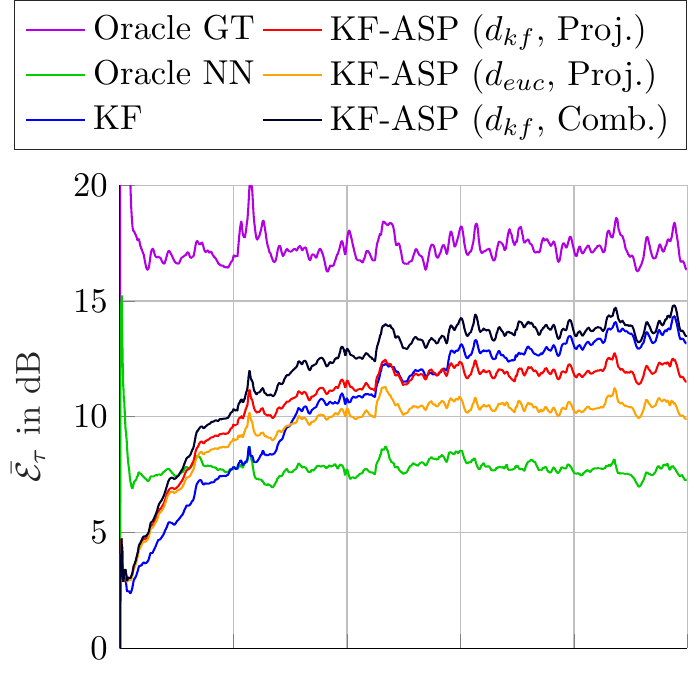}
	\hspace*{-.0cm}
	\includegraphics[]{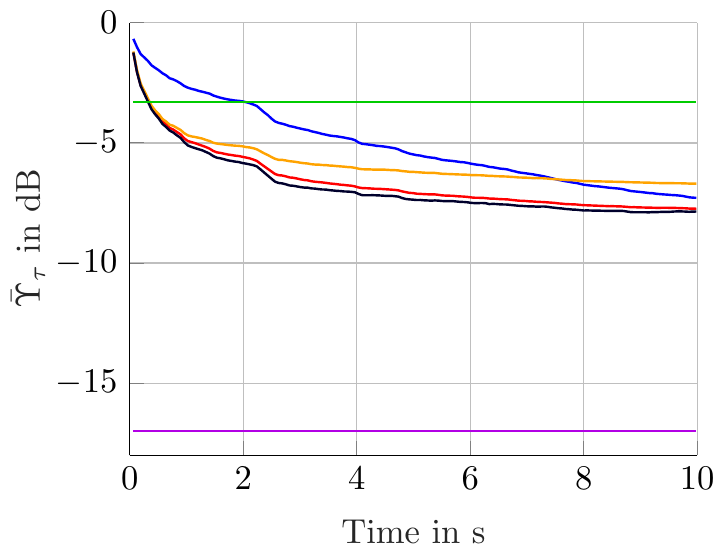}
	% \caption{rendering speech lr0 enr 0 dt Inf numTrainSamples 1000 \textcolor{red}{rather use a scenario with double talk here}}
		\caption{\commentTHc{Evaluation of the proposed KF-ASP-based algorithmic variants for a training data size of $\commentTHe{K}=1000$ samples. The loudspeakers play two independent speech signals ($\text{SNR}_{\text{wgn}}=0$ dB, $\text{SNR}_{\text{sp}}=\infty$ dB).}}
	%\caption{\commentTHc{Evaluation of affine subspace-based \ac{OSASI} algorithms for a training data size of $\commentTHc{K}=1000$ samples. The loudspeakers} play spatially uncorrelated speech signals. The noise-free observation vector is superimposed by a \ac{WGN} signal at an \ac{SNR} of $0$ dB.}% (\textcolor{red}{rather use a scenario with double talk here?})}
	\label{fig:red_sample_size}
\end{figure}

In the following experiments we evaluate the \commentTHe{proposed \commentTHe{KF-ASP} algorithm (cf.~Alg.~\ref{alg:prop_alg_descr})} for the previously \commentTHc{described} \ac{OSASI} scenario. As baseline we use the state-of-the-art \ac{KF} \cite{malik_recursive_2011} (cf.~Sec.~\ref{sec:kf_air_est}) with a state transition factor \makebox{$\commentTHd{A}=0.9999$}, a frame shift $R=512$, \commentTHd{a} filter length $L=512$ and recursive averaging factors $\lambda_{\text{W}}=0.9$ and $\lambda_{\text{N}}=0.5$. The state uncertainty matrix $\bP_\tau$ (cf.~Eq.~\eqref{eq:state_unc_cov_mat}) was initialized with an identity matrix scaled by $P_0 =0.01$. For the proposed algorithm we investigate three variants: \makebox{{KF-ASP ($d_{\text{kf}}$, Proj.)}}, \makebox{{KF-ASP ($d_{\text{kf}}$, Comb.)}} and \makebox{KF-ASP ($d_{\text{euc}}$, Proj.)}. Here, the variants including $d_{\text{kf}}$ use the probabilistic quadratic distance \eqref{eq:prob_dist_measure} whereas the last one, labeled by $d_{\text{euc}}$, uses the quadratic Euclidean distance \eqref{eq:euclidean_dist_metric} to compute the \commentTHe{$K_\tau=80$} \ac{NN} training samples for the update in time step $\tau$ (cf.~Sec.~\ref{sec:adSubProj}). The training data set is composed of $\commentTHe{K}=5000$ \commentTHc{spatially} uniformly distributed \acp{AIR}. Furthermore, we compare the effect of the convex combination-based enhancement (cf.~Sec.~\ref{sec:softSubProj}), labeled by \textit{Comb.}, in contrast to a hard projection, i.e., choosing $\alpha_{b,\tau,f}=1$ in Eq.~\eqref{eq:convex_combination}, \commentTHd{which is labeled by \textit{Proj.}} For the model prior (cf.~Eq.~\eqref{eq:model_prior}) $\beta_{\text{pr}}=5$ is chosen. The state uncertainty matrices of the \commentTHc{ASP}-based algorithms were initialized with \commentTHd{a scaling factor of} $P_0 =0.1$. Note that this higher initial state uncertainty could not be used in the \commentTHe{baseline} \ac{KF} as it led to divergence in our experiments. In addition, we evaluated two oracle baselines, i.e., {Oracle \commentTHc{GT}} and {Oracle NN}. Here, {Oracle \commentTHc{GT}} uses the first $L$ taps of the true \commentTHe{\acp{AIR} in} $\underline{\boldsymbol{w}}_\tau$ as estimate $\hat{\underline{\boldsymbol{w}}}_\tau$ and {Oracle NN} the training \commentTHe{sample} with the smallest squared Euclidean distance \eqref{eq:euclidean_dist_metric} to the true \ac{AIR} \commentTHe{vector}.

\commentTHc{Fig.~\ref{fig:wgn_exc} shows the average \ac{ERLE} $\bar{\mathcal{E}}_\tau$ and system mismatch $\bar{\Upsilon}_\tau$ for an excitation with spatially uncorrelated stationary \commentTHe{\ac{WGN}} input signal (\makebox{$\text{SNR}_{\text{wgn}}=0$ dB}, $\text{SNR}_{\text{sp}}=\infty$ dB).} We conclude from Fig.~\ref{fig:wgn_exc} that the proposed \commentTHe{KF-ASP} algorithms significantly \commentTHc{increase} the convergence rate of the baseline \ac{KF}. 
% \commentTHe{Furthermore, we observe that the proposed distance $d_{\text{kf}}$ performs similarly to the Euclidean distance $d_{\text{euc}}$. This is explained by the stochastic properties of the excitation and noise signals which do not result in a nonuniform frequency-dependent weighting in Eq.~\eqref{eq:non_diag_weight_kf_dist}. The} 
However, the steady-state performance of the \commentTHc{variants {KF-ASP} ($d_{\text{euc}}$, Proj.) and {KF-ASP} ($d_{\text{kf}}$, Proj.) which rely entirely on the hard projection, i.e., \commentTHd{the choice} $\alpha_{b,\tau,f}=1$ in the convex combination \eqref{eq:convex_combination}, are significantly worse than the baseline \ac{KF}.} This is due to \commentTHe{the imperfections} of the affine subspace model whose modeling capability \commentTHe{depends on the training sample size $K$ and the choice of the subspace dimension $D_\tau$ (cf.~Fig.~\ref{fig:subspace_system_mismatch}).} \commentTHe{In contrast, the soft projection-based \commentTHe{KF-ASP approach} (cf.~Sec.~\ref{sec:softSubProj}) achieves an increased steady-state performance in comparison the hard projection-based variants due to the convex combination (cf.~Eq.~\eqref{eq:convex_combination}) which takes the uncertainty of the model into account. Yet, the baseline \ac{KF} still shows a slightly improved steady-state performance which motivates an adaptive control of the prior covariance matrix \eqref{eq:model_prior}.}  \commentTHe{Furthermore, we observe that the probabilistic distance $d_{\text{kf}}$ performs similarly to the Euclidean distance $d_{\text{euc}}$. This is explained by the stochastic properties of the excitation and noise signals which do not result in a nonuniform frequency-dependent weighting in Eq.~\eqref{eq:non_diag_weight_kf_dist}.} %
\commentTHe{Finally, by inspecting the performance of the Oracle NN algorithm we conclude that the affine subspace models generalize better to \acp{AIR} in between the training data samples.}
% \commentTHe{Finally, we conclude that the affine subspace models generalize better to \acp{AIR} in between the training data samples than the {Oracle NN} algorithm which uses the \ac{NN} training \ac{AIR}.}
%by inspecting the {Oracle NN} algorithm, which uses the \ac{NN} training \ac{AIR},} we observe that the affine subspace models \commentTHc{generalize better} to \acp{AIR} in between the training data samples. 

A scenario with \commentTHe{the two loudspeakers playing independent speech excitation signals is shown in Fig.~\ref{fig:speech_exc} \makebox{($\text{SNR}_{\text{wgn}}=5$ dB, $\text{SNR}_{\text{sp}}=0$ dB)}.} 
%Here, we consider a mixture of a \ac{WGN} signal and an additional speech signal as \commentTHc{interferer}. The \ac{WGN} component is scaled according to an \ac{SNR} of $0$ dB and the interfering speech component to an \ac{SNR} of $5$ dB \commentTHc{w.r.t the noise-free observation}. 
Similarly as for the \ac{WGN} excitation the affine subspace projection greatly improves the convergence rate of the baseline \ac{KF}. However, in contrast to the \ac{WGN} excitation, the speech excitation benefits from the proposed probabilistic distance measure \commentTHc{$d_{\text{kf}}$} \eqref{eq:prob_dist_measure}. This is explained by the \commentTHc{nonuniform weighting in Eq.~\eqref{eq:non_diag_weight_kf_dist} which results from \commentTHe{the frequency-dependent} power of the speech excitation. Furthermore, the proposed KF-ASP algorithms show here an improved steady-state performance in comparison the \ac{KF} baseline and almost achieve the Oracle GT performance.} 
%\commentTHc{non-uniform frequency-dependency of the elements in the state uncertainty matrix} $\boldsymbol{P}_\tau$ which results from the speech excitation.

\commentTHc{We now} extend the previous system identification scenarios to a typical \ac{AEC} task in a teleconferencing setup \commentTHc{($\text{SNR}_{\text{wgn}}=10$ dB, $\text{SNR}_{\text{sp}}=\infty$ dB)}. Here, we assume that the loudspeaker array in the near-end room \commentTHc{plays} two microphone signals which are recorded in a distant far-end room \cite{enzner_acoustic_2014}. The far-end microphone signals are composed of the \commentTHc{reverberant} speech signals of two \commentTHe{spatially separated} far-end speakers with \commentTHc{disjoint activity}. Due to the high spatial correlation of the recorded speech signals, the system identification in the near-end room is complicated by the nonuniqueness problem \cite{sondhi_stereophonic_1995, sondhi_benesty_a_better_understanding}. Here, the near-end \ac{OSASI} \ac{AEC} algorithm  can only estimate a \commentTHc{nonunique cancellation filter} which often leads to severe performance drops whenever the \commentTHc{activity of the far-end speakers changes.} Fig.~\ref{fig:non_unique_evaluation} shows the average \ac{ERLE} $\bar{\mathcal{E}}_\tau$ and system mismatch $\bar{\Upsilon}_\tau$ computed from $50$ \commentTHc{trials for the experiment}. % We simulated two spatially separated far-end speakers whose activation flips after $5$ s. 
%Due to the non-uniqueness of the \ac{OSASI} problem the baseline \ac{KF} \commentTHc{obtains, despite the high \ac{ERLE}, a bad system mismatch.} 
\commentTHc{We conclude that, while the \ac{ERLE} is high, the baseline \ac{KF} does not converge towards the true \acp{AIR}. Due to the nonunique \commentTHd{filter} estimate the performance of the baseline \ac{KF} significantly drops after \commentTHe{the source activity switching at $t=5$ s}}. In contrast\commentTHd{, the KF-ASP-based algorithms achieve a much \commentTHe{better} average system mismatch $\bar{\Upsilon}_\tau$. This might be explained by the \commentTHe{projections keeping} the solution closer to the true \ac{AIR}.} Thus, there is no performance drop when the far-end \commentTHe{speakers flip}.
%
%}Thus, there is no performance drop when the far-end speaker changes its position.
% The far-end speakers are placed in a room of dimension $[x,y,z]$ m with a reverberation time of 

\commentTHc{Finally, we evaluate the effect of the training data size $\commentTHe{K}$. \commentTHd{To this end}, we simulate an additional scenario with spatially uncorrelated speech excitation ($\text{SNR}_{\text{wgn}}=0$ dB, $\text{SNR}_{\text{sp}}=\infty$ dB)}. However, in \commentTHc{contrast} to the previous experiments \commentTHc{($\commentTHe{K}=5000$)}, the affine subspace models are learned from a training data set including \commentTHe{only} $\commentTHe{K}=1000$ samples. The results shown in Fig.~\ref{fig:red_sample_size} suggest a rapidly \commentTHc{decreasing performance of the algorithms relying on hard projection onto affine subspaces.} \commentTHe{This is explained by the reduced number of training data samples $\commentTHe{K}$} \commentTHe{which only} allow to learn coarse \ac{AIR} models. In contrast\commentTHc{,} the soft projection-based approach, which does not enforce that all \ac{AIR} estimates are confined to the \commentTHc{subspace} model, still achieves an excellent steady-state performance in addition to the improved convergence rate.

\section{Summary and Outlook}
\label{sec:conclusion}
%\clearpage
%\newpage
%
In this paper we have introduced a \commentTHc{family of novel \ac{OSASI} algorithms which exhibit significantly faster convergence properties and improved steady-state performance in scenarios suffering from high-level interfering noise \commentTHe{and nonuniqueness of optimum filter estimates}.} The proposed \commentTHd{algorithms assume} that the variability of \acp{AIR} \commentTHc{in} an acoustic scene is confined to a \commentTHc{non-linear} manifold which can locally be approximated by an affine subspace. This allows to enhance an \ac{AF}-based \ac{AIR} estimate by projecting it onto the \commentTHc{learned} subspace.
\commentTHc{As future research we plan to develop improved choices for the prior covariance matrix and evaluate the effect of noisy training data samples on the learned \ac{AIR} models. }

\begin{backmatter}
\bibliographystyle{IEEEbib}
\bibliography{bmc_article}      % Bibliography file (usually '*.bib' )
% for author-year bibliography (bmc-mathphys or spbasic)
% a) write to bib file (bmc-mathphys only)
% @settings{label, options="nameyear"}

%\clearpage 
%\newpage
%
%\section*{TODO (8 - 10 pages)}
%
%%\begin{itemize}
%%	\item create experimental results as tikz and include into paper
%%	\item write down itemized fine structure of the sections
%%	\item write down continuous text 
%%\end{itemize}
%
%other items
%
%\begin{itemize}
%	\item if time allows you could consider a FD projection and use as a weighted reconstruction measure with low DFT bins having more power
%\end{itemize}

\end{backmatter}
\end{document}